\documentclass{article}

\usepackage[english]{babel}
\usepackage{caption}
\usepackage{subcaption}
\usepackage{siunitx}
\usepackage{lipsum}

\usepackage[table,xcdraw]{xcolor}
\usepackage[a4paper,top=2cm,bottom=2cm,left=3cm,right=3cm,marginparwidth=1.75cm]{geometry}

\usepackage{amsmath}
\usepackage{hyperref}
\usepackage{url}
\usepackage{graphicx}
\usepackage{floatrow}
\usepackage{authblk}

\usepackage{csquotes}
\usepackage[style=authoryear-ibid,backend=biber]{biblatex}
\usepackage[T1]{fontenc}
\usepackage{lipsum}

\title{Active-travel modelling: a methodological approach to networks for walking and cycling commuting analysis.}

\author{Ivann Schlosser}
\author{Valentina Marín Maureira}
\author{Richard Milton}
\author{Elsa Arcaute}
\author{Michael Batty}

\affil{Centre for Advanced Spatial Analysis, University College London, London, W1T 4TJ, UK}

\begin{document}
\maketitle
\normalsize
\begin{abstract}

Walking and cycling, commonly referred to as active travel, have become integral components of modern transport planning. Recently, there has been growing recognition of the substantial role that active travel can play in making cities more liveable, sustainable and healthy, as opposed to traditional vehicle-centred approaches. This shift in perspective has spurred interest in developing new data sets of varying resolution levels to represent, for instance, walking and cycling street networks. This has also led to the development of tailored computational tools and quantitative methods to model and analyse active travel flows.

In response to this surge in active travel-related data and methods, our study develops a methodological framework primarily focused on walking and cycling as modes of commuting. We explore commonly used data sources and tools for constructing and analysing walking and cycling networks, with a particular emphasis on distance as a key factor that influences, describes, and predicts commuting behaviour. Our ultimate aim is to investigate the role of different network distances in predicting active commuting flows. 

To achieve this, we analyse the flows in the constructed networks by looking at the detour index of shortest paths. We then use the Greater London Area as a case study, and construct a spatial interaction model to investigate the observed commuting patterns through the different networks. Our results highlight the differences between chosen data sets, the uneven spatial distribution of their performance throughout the city and its consequent effect on the spatial interaction model and prediction of walking and cycling commuting flows.

\end{abstract}

\section{Introduction}

Urban mobility, a key component of urban dynamics, intersects with various areas of urban life including economy, environment, and social structures. This critical aspect involves the movement of people and goods within urban areas, and its efficiency and sustainability directly influence the overall functioning of cities. Given its significance, it is crucial to acquire knowledge and understanding that help plan and develop improved mobility networks. This represents an ongoing challenge, as both transport methods and human mobility patterns continue to evolve.

An essential aspect of urban mobility is commuting, which is defined as the regular travel between residence and work places. Observing trends in commuting patterns can provide valuable insights into broader shifts in urban mobility. Nowadays, there is an increased focus on small-scale movement, primarily through cycling and walking. These modes of transportation are often grouped under the term "active-travel". Contemporary planning practices accentuate their importance for reasons of health, sustainability, and safety.

While active travel currently represents a moderate share of commuting methods in England, with about 10\% \parencite{office_for_national_statistics_2011_2017},
this figure is projected to increase. The rise in active travel not only signifies potential reductions in air and noise pollution due to less reliance on cars, but also implies urban redesign opportunities. The decrease in space required for parking and driving could allow for expanded green areas, and increased housing, leisure and commercial spaces \parencite{hidalgo_trillion_2020}. Given the current context of rising living and transport costs and global supply chain disruptions, active travel represents a practical and cost-effective transport alternative. Therefore, gaining a better understanding of the active travel flow patterns could help us adapt to these trends and meet the evolving needs of the population.

In this paper, we conduct a methodological analysis to study active travel. The study compares different walking and cycling networks to investigate how different models predict real-world flows. The findings of this research have tangible implications for policymakers and urban developers. By providing a more accurate understanding of active travel flows, the results can assist in the formulation of strategies that facilitate and promote walking and cycling.

The paper is structured into three main sections: Networks, Routing, and Modelling. In the first section, the focus is on constructing and comparing different walking and cycling networks from open data sources. In the second section different cost matrices are built based on shortest paths distances for each specific network using different origin-destination centroids. Upon examining the resulting detour index of various networks, we observe substantial disparities between the standard detour values and those that incorporate the real flows between locations. Even in the absence of ground knowledge of paths taken, relying on such metrics can prove helpful to construct efficient links between populations and job distributions. The final section integrates previous insights into a spatial interaction model, assessing the predictive capabilities of the different networks and routing parameters in relation to real walking and cycling flows. Finally, we combine these flows to model active travel commuting as a single phenomenon. We apply this method in London at the Middle Layer Super Output Area (MSOA) scale taking 2011 census data sets as reference for observed commuting flows.

The results show that choosing the appropriate network and distance measure can impact the accuracy of flow estimations, making it key to carefully consider network construction and analysis to ensure more accurate outputs.These findings not only bring attention to the differences between networks, but also stress how results diverge throughout the city, contingent upon the distribution and connectivity of networks as well as the uneven coverage of distinct data sources. 

\subsection{Walking and cycling}

Walking and cycling are active modes of transport because they require physical activity and are recognised for their potential health benefits and low carbon impact. Research has shown that they are associated with improved cardiovascular health, reduced risk of chronic disease, and better mental health outcomes \parencite{Saunders2013, Kelly2014, Oja2011}. They are considered sustainable modes of transportation due to their low carbon footprint and potential to promote transportation equity. In addition, they are cost-effective alternatives to traditional modes of transport. On the contrary, an increasing amount of literature is pointing at the fact that cars inside cities cause a series of problems that impact our health and well-being, such as noise and air pollution \parencite{fleury_geospatial_2021}, free space availability \parencite{hidalgo_trillion_2020}, congestion and heat island effects.

All those problems can be tackled with efficient public transport, active travel and policies. Thus, it is necessary to rethink and adapt infrastructure to better integrate them in the urban layout. Many of the worlds biggest cities are turning towards cycling as a sustainable and efficient mode of transport, incorporating new policies that facilitate the development of cycling infrastructure, and discouraging car driving. New York \footnote{\url{https://onenyc.cityofnewyork.us/wp-content/uploads/2019/05/OneNYC-2050-Efficient-Mobility.pdf}}
, Chicago\footnote{\url{https://www.chicago.gov/content/dam/city/depts/cdot/CDOT\%20Projects/Strategic_Plan/Strategic_Plan_for_Transportation21.pdf}}
, San Francisco \footnote{\url{https://sfgov.org/sfplanningarchive/ftp/General_Plan/I4_Transportation.htm}}
, Boston \footnote{\url{https://www.boston.gov/sites/default/files/document-file-06-2018/ib2030_book_spreads-transportation.pdf}}
, Singapore \footnote{\url{https://www.ura.gov.sg/Corporate/Planning/Master-Plan/Themes/Convenient-and-Sustainable-Mobility}}
and others mention cycling and walking as central elements of their current and future public transport policies in their latest master plans.

Walking and cycling share many similarities as active travel modes yet also exhibit some distinctions. While cycling is typically characterised as faster and more suitable for longer journeys, it necessitates specialised equipment and could present more safety risks due to the increased vulnerability of cyclists on the road. Walking is more accessible, exempt from equipment requirements, and generally considered safer. These modes differ substantially in terms of their typical movement speeds. For walking, speeds of around $V_w = 1.3\unit{m.s^{-1}} (4.6 \unit{km.h^{-1}})$ \parencite{HealthWalk,Hamer238} 
are usually assumed, while for cycling the value are usually of about $V_c = 4.2 \unit{m.s^{-1}} (14 \unit{km.h^{-1}})$ \parencite{romanillos_pulse_2018}. This results in significant differences in the area that can be covered within a given time, with cycling being a more efficient way to cover longer distances while walking may be better suited for short journeys.

\subsection{Commuting and recreational trips}
As with other modes of transport, it is common to distinguish trips by purpose, commuting on one hand, and leisure on the other. The two types of travel have varying influencing factors, and require different approaches for analysis and specific planning and policy interventions \parencite{Krizek2009}. 

Factors such as a high-quality built environment, urban amenities, infrastructure, land use mix, and street scenery can impact the decision to walk or cycle for both recreational and commuting trips \parencite{Schoner2014, Cervero2019}, but also factors like safe and diverse route options and the availability of destinations within a reasonable walking or cycling distance. Specifically for cycling, there is consistent evidence that the quality, extent, and connectivity of a road or cycle network have a positive correlation with cycling uptake \parencite{Cervero2019,dill_bicycle_2003}. Under the context of walking and cycling for commuting, time and distance are crucial factors. Recreational trips are typically less frequent, more flexible, and do not prioritise travel time/distance efficiency. Conversely, commuting trips are done regularly, with time constraints being a significant factor in route choice. As a result, commuting trips tend to prioritise distance and be less responsive to other variables \parencite{Broach2011}. 

With distance being identified as a major barrier to active travel, it is essential for predicting and explaining travel behaviour. As such, analysing shortest paths within cycling and walking routes provide a useful benchmark for identifying optimal commuting routes. While shortest paths may not always align with observed route choices, empirical evidence suggests that people generally do not deviate significantly from the shortest distance route, with most trips being less than 10$\%$ longer than the optimal path \parencite{Winters2010, Mahfouz2021}.

\section{Networks}

This section considers the aspects directly related to downloading and manipulating open data to construct road networks, modelled as graphs with positive edge lengths. Using different types of networks can further on allow us to build varying distance matrices to examine the potential active travel between locations. 

When it comes to calculating the shortest paths between origins and destinations, there are different distance metrics that can be used. On top of considering euclidean distance between locations, it is important to account for the actual distance travelled on the road network between two points. Doing so allows to build a more accurate understanding of the spatial distribution of trip lengths originating from or reaching a place, thus uncovering some connectivity aspects, local network properties and their impact on the magnitude of active travel flows. 

\subsection{Road networks}
Networks representing physical roads have already received a fair amount of attention in the literature and their properties have been observed across geography and even time \parencite{strano_elementary_2012,masucci_exploring_2014,barthelemy_spatial_2011}, to study economic indicators \parencite{porta2009street,porta2012street,law2013measuring,Piovani2017,}, to perform comparative analysis between cities \parencite{hillier2012normalising,strano2013urban,}, to examine social and demographic indicators \parencite{vaughan2007spatial,molinero2015fractured}, and asses accessibility to land-uses \parencite{freiria2015multiscale,novak2014link,}, to name a few.  In the context of network science, street networks are commonly represented using a directed or undirected primal and planar graph. In primal graphs, a road network is modelled using links to represent the street segments and nodes to represent street junctions \parencite{porta2006network, marshall2018street,}. While planar means that the network is constrained to a two-dimensional space.   

The level of detail required to construct street networks for graph analysis can vary depending on the specific analysis objectives, which may have different effects on the outcomes \parencite{marshall2018street,barthelemy_spatial_2011,jungGravityModelKorean2008}. For analyses at a very local scale, a more intricate street network is typically used, which is constructed by taking into account pedestrian-specific infrastructure like crosswalks and sidewalks \parencite{palominos_examining_2022,Rhoads2020, thompson_sargoni_neighbourhood-level_2023}. In studies focused on spatial configuration and human behaviour, such as those using the Space Syntax theory, street networks are constructed using centre lines that represent angular changes \parencite{Penn2003}. In this paper, we aim to use open data sources and construct an undirected primal and planar graph with an optimal level of resolution for active travel routing and modelling. To do so, we use the two most detailed available representations of the road network currently available for the UK. 

First, England's official provider of digitised infrastructure data, Ordnance Survey \footnote{\href{https://www.ordnancesurvey.co.uk/business-government/products/open-map-roads}{https://www.ordnancesurvey.co.uk/business-government/products/open-map-roads}} covering England, Wales and Scotland, is used. Second, OpenStreetMap (OSM) 
, an open source platform aiming at digitising as much of the physical world as possible. It does so through user contributed map features that have a detailed tagging system associated to them. The high quality of the road networks data makes it highly used in modern research on the built environment. Among the advantages of using OSM are transferability between areas of observation and scalability, making any analysis scalable to the whole planet. It allows large scale road network analysis \parencite{gallotti_unraveling_2021,louf_typology_2014,barthelemy_paths_2017,lovelace_propensity_2017}, going beyond the borders of a single country \parencite{boeing_street_2021, liu_generalized_2021}, while maintaining confidence in the quality of the data.

\begin{figure}[!ht]
\begin{center}
\includegraphics[width=15cm]{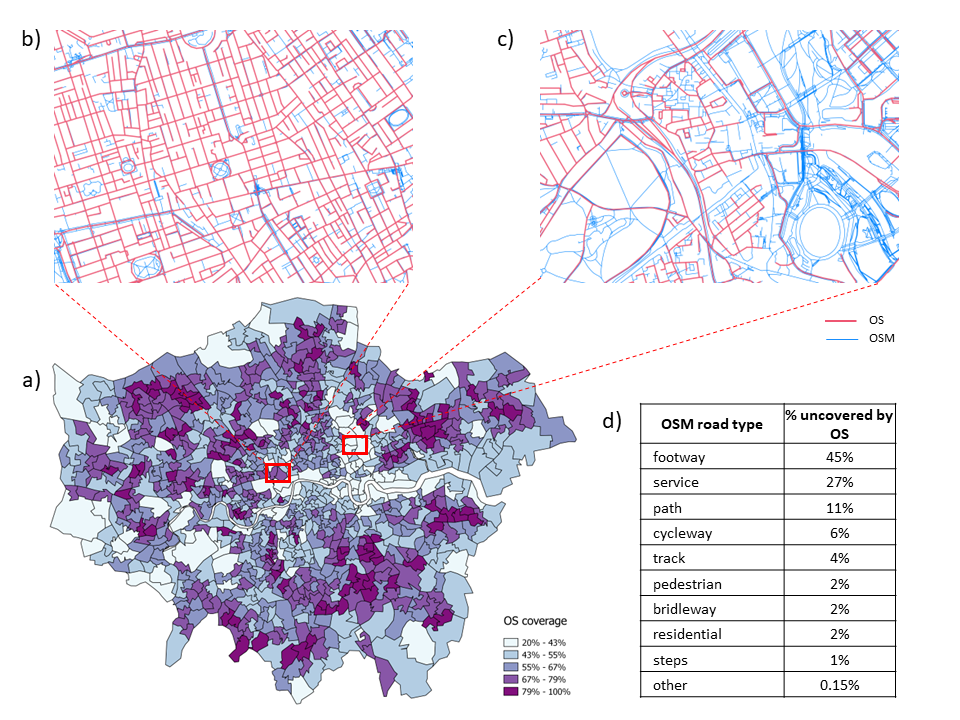}
\caption{\textbf{OS and OSM road network coverage in London, UK.} a) Percentage of OS over OSM road coverage, based on road linear metres per MSOA level. b) and c) Overlapping of OS and OSM networks showing differences in coverage depending on the urban structure of London areas, Marylebone and Stratford respectively. Table in d) shows the percentage of linear metres from roads on OSM that are not covered by the OS network. Percentages are organised by road types.}
\label{OSvsOSM}
\end{center}
\end{figure}

The two sources used present some differences that need to be put in context, as evidenced by the comparison in figure \ref{OSvsOSM}. Although both data sets could differ in terms of geographical extent, geometrical shape and attribute accuracy of segments, this comparative study focuses mainly on the completeness of the data sets. We examine to what extend their coverage is suited for the analysis of pedestrian and cycling networks. OS Open Roads  considers exclusively roads that are used by cars, while OSM represents a wider range. On figure \ref{OSvsOSM} one can see the two data sets overlayed on two areas: in Fitzrovia (b) and Stratford (c). The blue network being OSM and red being OS. In b), most of the roads can be taken by car, the only differences can be seen in pedestrian paths crossing squares and parks. But when looking at the area of Stratford, the differences become more striking. In this case, the OSM network presents a much greater set of roads, which are primarily dedicated to pedestrians and cyclists, including those around and inside the mall and the Olympic park. Further inspection shows that the difference of road linear metres coverage for both networks is unevenly distributed across MSOAs (Figure \ref{OSvsOSM}a). Knowing in advance that OS is a more high-level representation of roads, the higher coverage percentage of OSM is not a surprise. However, the high difference between MSOAs is a fact to consider when analysing the results in areas with different urban structures. We then examined whether the difference was consistent across street types, so we measured the percentage that are not covered by the OS dataset. Results show that streets classified as footways, service, paths and cycleways in OSM are under covered by OS (Figure \ref{OSvsOSM}d).

\subsection{Network setup}

This section introduces some of the commonly used open source tools to download and construct a road network. OSM data can be downloaded from a number of sources, including the OpenStreetMap website and dedicated download servers like Geofabrik \footnote{\url{https://www.geofabrik.de}}, which help download raw OSM data in different formats as files with multiple layers of features. Optimal access to this data can be achieved by using  many software packages across different programming languages. We mainly consider R and python packages as they are widely used in the community. With each being designed for different purposes, the format and attributes within the data can vary. We considered a non exhaustive list of packages and the main criteria used to select the appropriate one are shown in table \ref{osm_packages}. \textbf{OSMnx}\footnote{\url{https://OSMnx.readthedocs.io/en/stable/}} \parencite{boeing_OSMnx_2017} is a python package,\textbf{osmdata} \footnote{\url{https://docs.ropensci.org/osmdata/}} \parencite{osmdata} and \textbf{osmextract} \footnote{\url{https://ropensci.github.io/osmextract/index.html}}
\parencite{osmextract} are R packages. We examine and compare different sources for downloading and working with OSM road networks to evaluate which is the most suitable considering the requirements of our analysis (Table \ref{osm_packages}). Therefore, we will be using OSMnx as the source of OSM data.

\begin{table}[h!]
\centering
\resizebox{\textwidth}{!}{%
\begin{tabular}{
l 
c 
c 
c 
c }
{ \textbf{}} & { \textbf{osmdata}} & { \textbf{osmextract}} & { \textbf{Geofabrik}} & { \textbf{OSMnx}} \\
\textit{Big data} & no                                      & yes                                        & yes                                       & yes                                   \\
\textit{Optimal for building graphs}      & no                                      & no                                         & no                                        & yes                                   \\
\textit{High details} & yes                                     & yes                                        & yes                                        & no                                   
\end{tabular}%
}

\caption{Overview of packages for working with \textbf{OSM} road networks. This list is non exhaustive and contains the options considered \parencite{Mahfouz2021,boeing_morphology_2018,costa_circuity_2021} as well as the main criteria for selection. The first category, \textit{Big data} implies that it is possible to download region and country scale data sets. \textit{Optimal for building graphs} means that the downloaded data is structured in a way that allows to build graphs without further engineering. Usually, it would be divided into two data sets, one containing nodes with their id, coordinates and additional optional attributes. The other data set contains edge ids, origin node id, destination node id and extra optional arguments such as length. Finally, \textit{High details} means that several columns of OSM key$\sim$tag values are present additionally to the main classification corresponding to the highway 
key. Additionally, some even more powerful command line tools are available, and would be useful for a bigger scale of analysis, but are omitted from this work.}
\label{osm_packages}
\end{table}

\begin{figure}[!ht]
\begin{center}
\includegraphics[width=15cm]{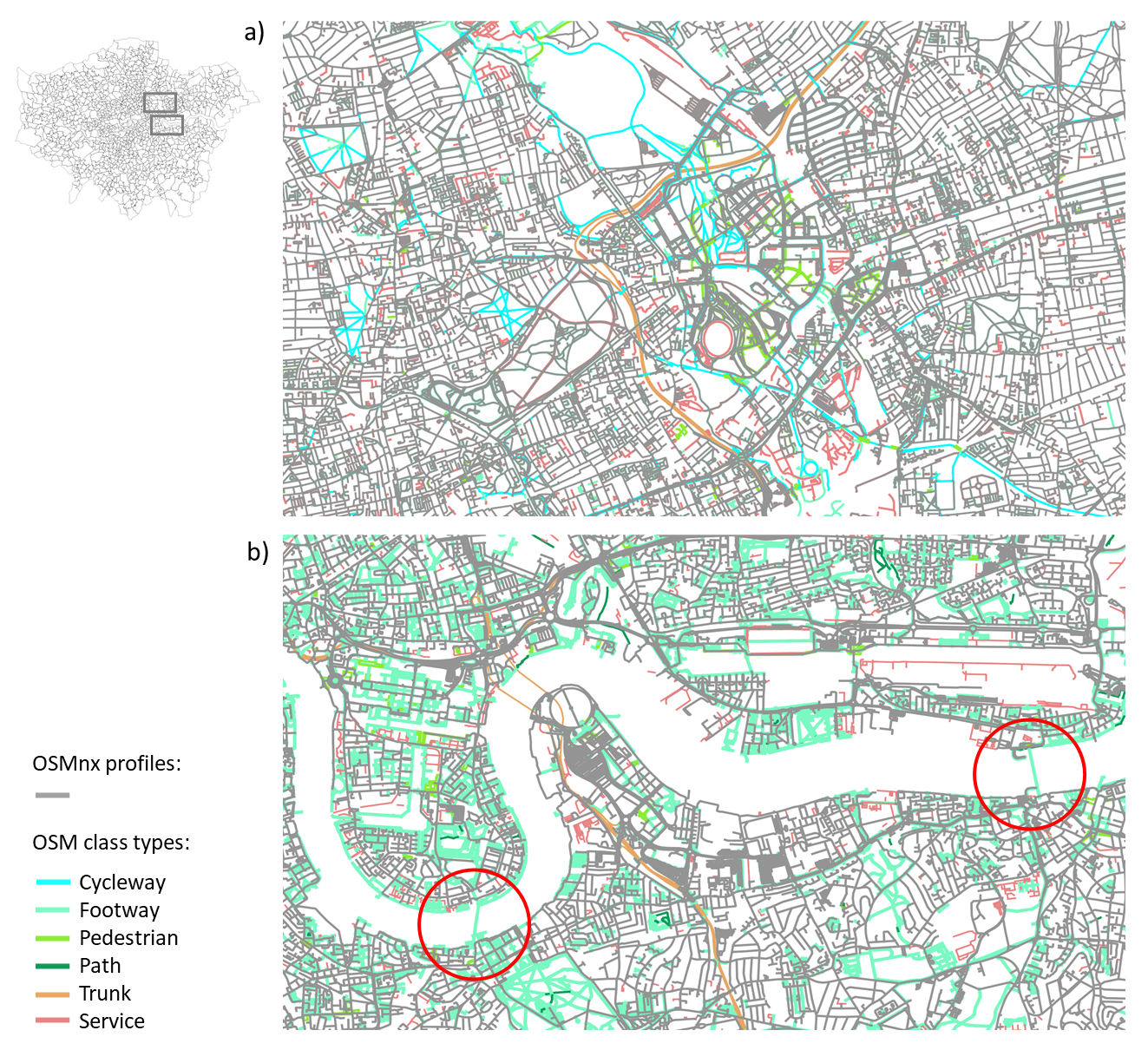}
\caption{\textbf{OSMnx walk and cycle profiles overlapping the OSM full network.} The OSM full network in different colours showing the class types that are filtered out in the walk and cycle network. a) In grey the OSMnx walk profile on top of the full network. In cian The Greenway footpath running from Victoria park to the southeast, and classified as cycleway. Other footpaths within parks are also classified as cycle ways only. b) In grey the OSMnx cycle profile, in green, the Woolwich and Greenwich foot tunnels highlighted. Given that these tunnels are classified as footways in the OSM network, they are filtered and not considered within the cycle profile.}
\label{OSMnx_profiles}
\end{center}
\end{figure}

As streets have varying attributes and serve distinct purposes, network modelling for specific modes of transportation, such as buses or private cars, only incorporates streets suitable for those modes. However, when it comes to walking and cycling networks, the definition is less precise as pedestrians and cyclists can move more freely throughout the city. This has promoted the interest for creating alternative network profiles using the OSM network as a base. Therefore, additionally to downloading the desired networks, some packages offer the functionality to filter the roads based on a specific mode of transport, like walking, cycling or driving. The OSMnx package in python provides profiles for different transport modes by filtering out streets based on the tag keys. The dodgr \footnote{\url{https://atfutures.github.io/dodgr/}} \parencite{dodgr} package in R creates alternative profiles by weighting the segments according to different attributes such as speed, preferences and restrictions. 
Moreover, since both profiles rely on the accuracy of the attribute tags given to each street when it is mapped on the OSM network, we first examined whether some of the streets were miss-classified resulting in the inappropriate exclusion or inclusion of street segments (Figure \ref{OSMnx_profiles}).

By looking at figure \ref{OSMnx_profiles}, the streets that were filtered out (in colors) from each profile (in gray) can be identified. For instance, in the case of walking (a), some routes along parks and The Greenway footpath, which runs from Victoria Park to the southeast are excluded from the network as are classified as cycleways in the OSM dataset. A similar case is that of (b), where the tunnels that cross the Thames are excluded from the cycling network as they are classified as footways. As a result, connectivity costs are increased, which may  affect global and local patterns, such as the connectivity within East London or between the north and south areas of the river.

Based on these initial observations, different network profiles are used to create distinct OD cost matrices between MSOAs, including those provided by popular packages, and one where minimal filtering was done (motorways removed as forbidden to cycle or walk on) to serve as a base model for our tests. The Ordnance Survey (OS) data is also used as it is the official and open source reference for the road network in the UK.

\section{Routing} \label{routing}

The next step in developing our analysis consists in using the networks for routing origins and destinations. In the scope of this work, we will only look at the distances that separate OD pairs in the network by using shortest paths algorithms. To accomplish this, we compared a set  of commonly used packages for computing shortest paths. We then examined various approaches to determine origin-destination points for commuting, particularly when data is aggregated at geographical units, such as the MSOA level in this case. As a result of this procedure, different cost matrices are created, one for each network, which will serve as inputs for the following section to obtain the prediction of pedestrian and cycling commuting flows.

\subsection{Shortest paths}

We consider the distance between an origin and destination in a specified network to be the length of the shortest path that connects them. An alternative approach that can be used and is already implemented with certain packages is to consider time. An extra layer of sophistication can also be added by attributing additional weight factors to each link of the network based on its classification and the willingness of cyclists or pedestrians to use it. For example, a commuter might have a tendency to use a smaller, quieter route of greater length rather than a shorter, but busy and risky one. For the purposes of our work, we prioritise distance as the most critical factor for commuting decision-making, which also provides an approach that is easy to implement for a large scale analysis and can be further developed with more variables in future works. The other methods could be useful to learn more about path choice of commuters, who on the one hand think about their safety and comfort, but on the other aim to minimise their trip distance or duration. The significance of other factors may be challenging to validate and is left out of the scope of this work.

\begin{table}[]
\centering
 \caption{Mean and median times in seconds to compute a distance matrix between 50 and 100 randomly chosen locations. Different types of heaps are tested for dodgr and two different algorithms for cppRouting, as well as multiple core run when this functionality is provided. \textbf{sf\_networks} \parencite{sfnetworks} is a package that, together with \textbf{tidy\_graph} \parencite{tidygraph} relies on \textbf{igraph} \parencite{igraph} for routing. Tests are done with the \textbf{microbenchmark} \parencite{microbench} package in R. The hardware used is a 2022 M2 MacBookPro with 16 GB RAM. }
\label{tab:package_performance}
\begin{tabular}{cc|cc|cc|}
\cline{3-6}
                                            &                 & \multicolumn{2}{c|}{\textit{1 thread, 50x50 OD}}     & \multicolumn{2}{c|}{\textit{4 threads 100x100 OD}}   \\ \hline
\multicolumn{1}{|c|}{\textbf{package}}      & \textbf{option} & \multicolumn{1}{c|}{\textbf{mean}} & \textbf{median} & \multicolumn{1}{c|}{\textbf{mean}} & \textbf{median} \\ \hline
\multicolumn{1}{|c|}{\textbf{tidygraph}}    & -               & 5.13                               & 5.13            & -                                  & -               \\ \cline{1-2}
\multicolumn{1}{|c|}{\textbf{sf\_networks}} & -               & 4.10                               & 4.06            & -                                  & -               \\ \cline{1-2}
\multicolumn{1}{|c|}{\textbf{dodgr}}        & fheap           & 24.12                              & 22.92           & 18.66                              & 17.81           \\ \cline{1-2}
\multicolumn{1}{|c|}{\textbf{dodgr}}        & bheap           & 16.87                              & 16.89           & 13.91                              & 13.90           \\ \cline{1-2}
\multicolumn{1}{|c|}{\textbf{dodgr}}        & triheap         & 26.96                              & 26.61           & 20.54                              & 19.79           \\ \cline{1-2}
\multicolumn{1}{|c|}{\textbf{cppRouting}}   & phast           & 2.40                               & 2.41            & 1.37                               & 1.37            \\ \cline{1-2}
\multicolumn{1}{|c|}{\textbf{cppRouting}}   & mch             & 0.35                               & 0.34            & 0.37                               & 0.37            \\ \hline
\end{tabular}
\end{table}

Based on the outputs from the benchmark shown in table \ref{tab:package_performance}, we use the cppRouting package in R to perform the one-to-one and many-to-many calculations and to build distance matrices. This package implements fast, multi-threaded processes, relying on the Dijkstra \parencite{stein_introduction_2009}, A-star and PHAST algorithms as well as a contraction hierarchy optimisation which greatly reduces computational times on big graphs. 

\begin{figure}[!h]
\begin{center}
\includegraphics[width=12cm]{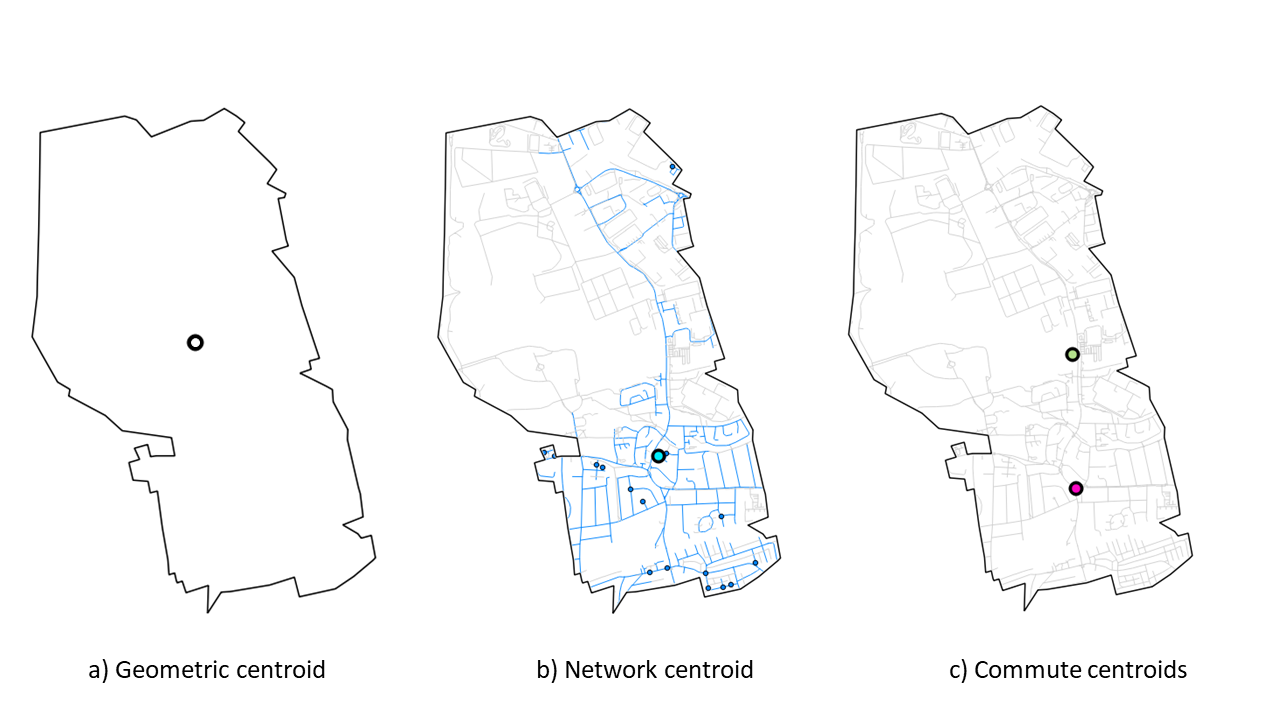}
\caption{\textbf{Origin-destination centroids (Beddington MSOA illustrated here).} a)The geometric centroids from every MSOA polygon. b) The network centroid based on the distribution of the street network within the MSOA boundary by taking the centre of mass of a sample of road intersections between ``living street", ``primary", ``secondary", ``tertiary", ``trunk" and ``residential" OSM road types (in blue). c) The "commute" centroids, based on the population distribution (origin) in pink and job location (destination) in yellow.}
\label{centroids_image}
\end{center}
\end{figure}

\subsection{Selecting origins and destinations}\label{what_to_what}

The next aspect of routing consists in solving the what to what problem, namely determining the actual locations to use as the origin and destination. This step is particularly important when the data is aggregated by relatively big units of space (MSOA). For smaller units, like output areas (OA), this question might not be as relevant, because they cover only the area of a few houses, and thus a regular centroid will not bear significant difference from any other node within an area. However, for larger areas that span about $1\unit{km}$ and include multiple streets, overlap parks, rivers and rails, this can become significant, as is demonstrated later on. Although it seems appropriate to choose the smallest available resolution, breaking down the system into small sub units comes at a great cost when running any model. 
Aiming to develop a methodology that could further on be scaled up to the country level, the MSOAs were used, with 983 in London. 

We further consider 3 types of MSOA centroids that are illustrated in figure \ref{centroids_image}. The first one is the \textit{geometric} centroid, that is quite commonly used in research \parencite{bassolas_hierarchical_2019,zhong_detecting_2014}. The second one is referred to as \textit{network} centroid and is the centroid of a sample of nodes that lie within residential roads, this allows to exclude parks, industrial or shopping areas. The third one is a tuple of nodes. The first being the population weighted centroid \footnote{\url{https://geoportal.statistics.gov.uk/datasets/ons::msoa-dec-2011-population-weighted-centroids-in-england-and-wales/about}} that accounts for the distribution of residents in space, and the other being the workplace weighted centroid, that is obtained using the definition of workplace zones\footnote{\url{https://geoportal.statistics.gov.uk/datasets/ons::workplace-zones-december-2011-population-weighted-centroids-in-england-and-wales/about}} in the UK statistics office. We refer to this pair as the \textit{commute} centroids. For each centroid point, the nearest network node is assigned and the distance between them is added to the final matrices. 

\subsection{Flows data}

In the following, the methods and data described earlier will combined with flows data to develop a metric that helps understand the commuting patterns of active travellers.

The commuting data in this work comes from the 2011 UK census and is accessed through the governmental data portal (See details in the appendix). It is available in its raw format for research purposes to individuals linked to a university. The flows contain information on the origin area, destination area, and method of travel, for more details on the variables refer to the online material\footnote{\url{https://github.com/ischlo/QUANT_at}}. Different levels of spatial units are used in the UK census, and this study relies on MSOAs, which are constructed to contain around 8000 individuals each. In this work, only the data for cycling and walking is used for the flows. Additionally, the flows on distances beyond 15km are considered as outliers or collection errors and are excluded. 

\subsection{Detour index} 

When measuring the performance of networks in terms of distance, as in the context of transport studies, it is useful to look at a measure called the detour index, also referred to as circuity \parencite{barthelemy_spatial_2011}, which can take different forms if we consider a whole network or a node, or a specific route:

\begin{equation}
    \delta_{ij,k} = \frac{l_{ij,k}}{d_{ij}}
\end{equation}

where $l_{ij,k}$ is the distance in the network $k$ between location $i$ and $j$, and $d_{ij}$ is the Euclidean distance between them. If aggregating on an origin node and considering a set of destination nodes, which is not necessarily all the nodes in the network: 

\begin{equation}
    \delta_{i,k} = \frac{1}{N_S-1} \sum_{j\in S,i\neq j} \delta_{ij,k}
\end{equation}

which gives the average detour index of the routes from node $i$ to all other nodes of interest in set $S$ for a specified network $k$.
And, for the whole network other the set of nodes $S$:

\begin{equation}
    \delta_{k} = \frac{1}{N_S(N_S-1)} \sum_{ij\in S,,i\neq j} \delta_{ij,k}
\end{equation}

where $N_S$ is the number of nodes in set $S$. The detour index is greater or equal to 1, and the more efficient, or straight, a network path is the closer the value will be to 1. Typical values are between $1.1$ and $1.5$ and values over 2 show a poor performance  \parencite{levinson_minimum_2009,yang_universal_2018}. This index has been linked to important aspects of the network such as accessibility and efficiency. In \parencite{costa_circuity_2021}, the detour index was found to decrease with time in cycling networks. This indicates a tendency of the network to ameliorate it's connectivity, by growing new links or changing the routing possibilities. In \parencite{levinson_minimum_2009} it was observed that users of the network tend to locate in places that minimise detour for their commuting trip compared to a random selection of origin-destination pairs. The authors observe different values of average detour for random selections of OD nodes compared to observed OD flows, indicating that the efficiency of the network is better understood through the mobility patterns occurring in it. Hence, the consideration of a subset $S$ of the nodes involved in observed OD flows, in our case the MSOA with flows between them, is relevant to better understand commuting trip distributions in the network. 

In this section we took the different street networks and consider the geometric centroids as origin-destination points for a straightforward comparison. 

\begin{figure}[!ht]
\begin{center}
\includegraphics[width=16cm]{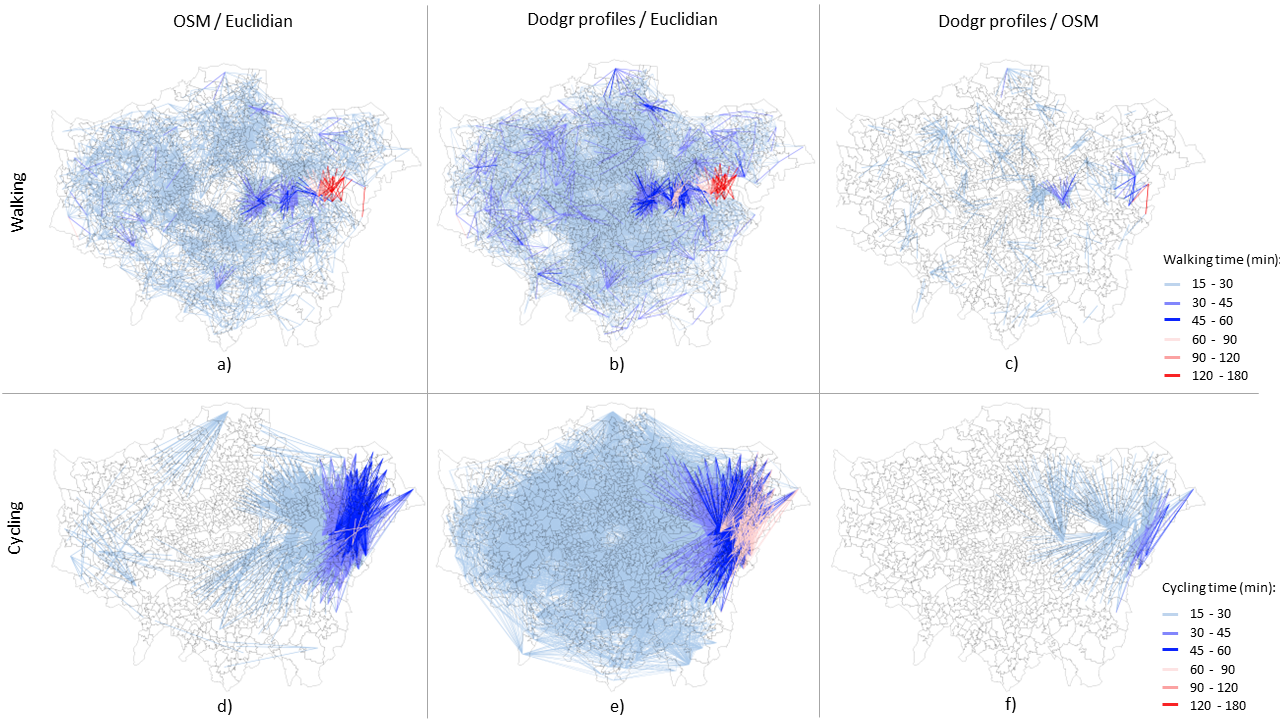}
\caption{\textbf{Detour given walking and cycling times when comparing shortest paths from the OSM and dodgr networks with the euclidian distances between MSOAs.} We converted distance into time cost using 1.3 m/s and 4.2 m/s as the walking and cycling speeds, respectively. We excluded trips wich would take longer than 1 hour as a maximum threshold for commuting by these modes. We visualised differences in time longer than 15 minutes as a way to highlight detours that would have a greater impact on the decision to use active travel}
\label{cycling_walking_times}
\end{center}
\end{figure}

We plot the detour distance values in Figure \ref{cycling_walking_times}, translated into walking and cycling times, to make the comparison of both modes easier. Differences in the cost of the shortest paths between MSOAs are considerable across space. Results are highly determined by the uneven distribution of the street network provision and network impedance. Due to the lack of crossings in the east, for example, MSOAs at each side of the Thames present the higher detour values with commuting times reaching a difference of up to 3 hours. Higher differences can be found in MSOAs with large green areas, large block sizes and other barriers like train lines. Based on the results, when comparing  the different commuting times it is important to consider that the differences between the network distances are not evenly distributed in space, and that the selection of one or the other may affect the results of future analyses in specific areas of the city.

\begin{figure}[!h]
    \begin{subfigure}{0.49\textwidth}
        \includegraphics[width=\linewidth]{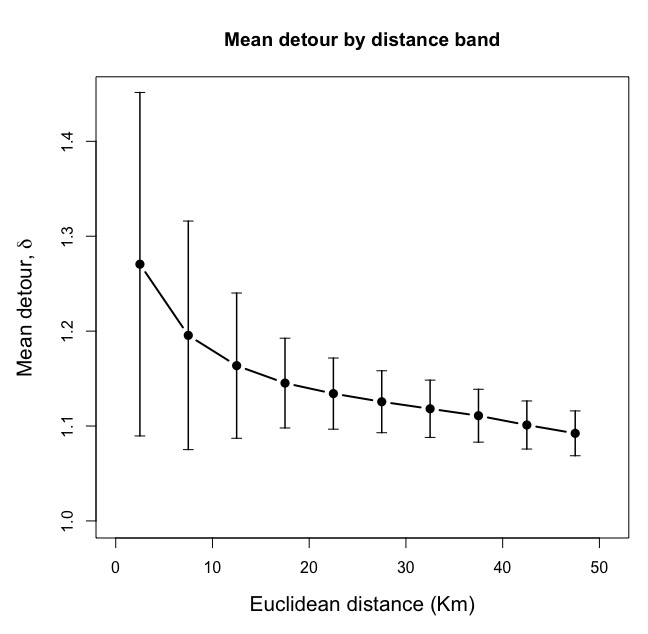}%
        \caption{}
        \label{detour_dist_band}
    \end{subfigure}
    \hfill
    \begin{subfigure}{0.49\textwidth}
        \includegraphics[width=\linewidth]{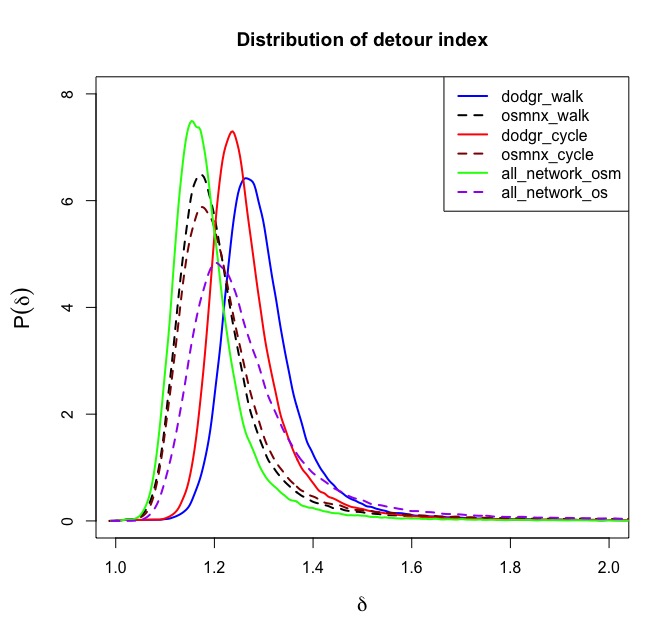}
        \caption{}
        \label{detour_distr}
    \end{subfigure}
    
    \begin{subfigure}{0.49\textwidth}
        \includegraphics[width=\linewidth]{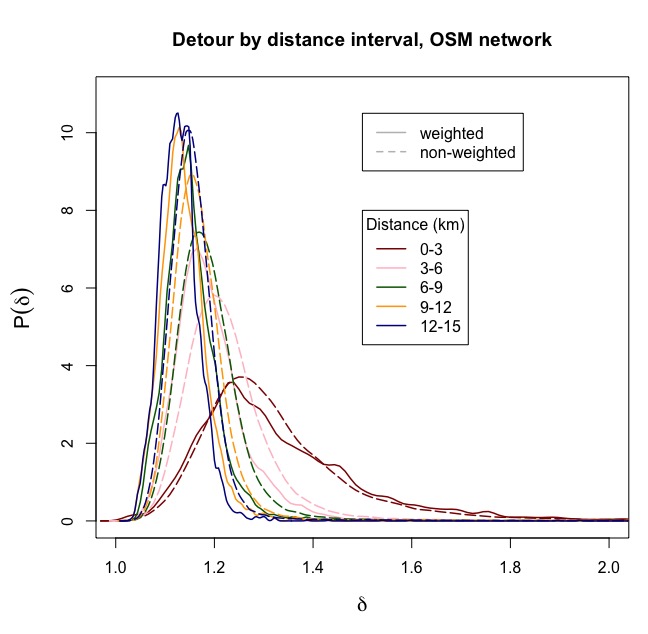}%
        \caption{}
        \label{detour_band_osm}
    \end{subfigure}
    \hfill
    \begin{subfigure}{0.49\textwidth}
        \includegraphics[width=\linewidth]{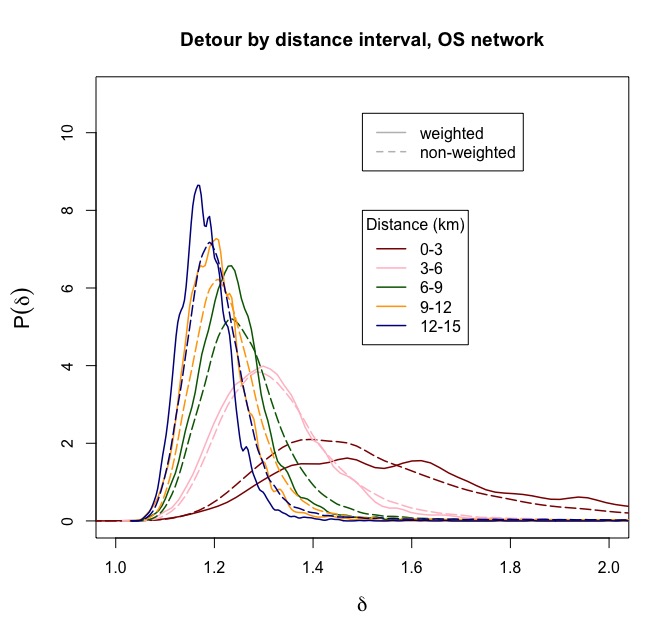 }%
        \caption{}
        \label{detour_band_os}
    \end{subfigure}

    \caption{\textbf{The detour index} is the ratio of the distance between two points in the network chosen and the shortest, Euclidean distance, $\delta^k_{ij} = \frac{l_{ij,k}}{d_{ij}}$, where $l_{ij,k}$ is the distance in the network $k$, and $d_{ij}$ is the Euclidean distance between those points. Clearly, $\delta \in [1,\infty [$. One can see that the kind of routing (fig \ref{detour_distr}) proposed by the $\boldsymbol{dodgr}$ package results in longer shortest paths due to the weighting that favours smaller roads to bigger, busier ones for cyclists and pedestrians. The network that approaches the most the theoretical case of $\frac{\sum_{i,j = 1}^N\delta^k_{ij}}{N(N-1)} = \bar{\delta^k}$ = 1, is the OSM full network, which contains short cuts and links not included elsewhere. The Ordnance Survey network (dashed purple line) is the least efficient of the networks without custom weighting profiles. On \ref{detour_dist_band}, one can observe the decreasing average detour and shrinking standard deviation as the Euclidean distance between locations increases. On figures \ref{detour_band_osm} and \ref{detour_band_os}, the detour distribution, weighted (solid line) and non weighted (dashed line) by commuter flow for different distance intervals of Euclidean distance of trips. The table in the appendix, presents the same observation for finer distance intervals, removing the effects of the growing number of destinations with distance.}
    \label{detour_fig}
\end{figure}

\begin{table}[]
\centering
\begin{tabular}{llllllll}
\multicolumn{2}{c}{\textit{d (km)}}                  & \multicolumn{3}{c}{\textit{OS}}                                                               & \multicolumn{3}{c}{\textit{OSM}}                                                             \\
\multicolumn{1}{c}{min}   & \multicolumn{1}{c|}{max} & \multicolumn{1}{c}{$\delta_c$} & \multicolumn{1}{c}{$\delta_n$} & \multicolumn{1}{c|}{t-test} & \multicolumn{1}{c}{$\delta_c$} & \multicolumn{1}{c}{$\delta_n$} & \multicolumn{1}{c}{t-test} \\ \hline
\multicolumn{1}{|l|}{0.0} & \multicolumn{1}{l|}{0.6} & \multicolumn{1}{l|}{2.113}     & \multicolumn{1}{l|}{2.108}     & \multicolumn{1}{l|}{0.1}    & \multicolumn{1}{l|}{1.498}     & \multicolumn{1}{l|}{1.518}     & \multicolumn{1}{l|}{-0.8}  \\
\multicolumn{1}{|l|}{0.6} & \multicolumn{1}{l|}{0.9} & \multicolumn{1}{l|}{1.931}     & \multicolumn{1}{l|}{1.914}     & \multicolumn{1}{l|}{0.9}    & \multicolumn{1}{l|}{1.454}     & \multicolumn{1}{l|}{1.471}     & \multicolumn{1}{l|}{-1.6}  \\
\multicolumn{1}{|l|}{0.9} & \multicolumn{1}{l|}{1.2} & \multicolumn{1}{l|}{1.780}     & \multicolumn{1}{l|}{1.791}     & \multicolumn{1}{l|}{-0.7}   & \multicolumn{1}{l|}{1.384}     & \multicolumn{1}{l|}{1.423}     & \multicolumn{1}{l|}{-4.9}  \\
\multicolumn{1}{|l|}{1.2} & \multicolumn{1}{l|}{1.5} & \multicolumn{1}{l|}{1.666}     & \multicolumn{1}{l|}{1.700}     & \multicolumn{1}{l|}{-3.0}   & \multicolumn{1}{l|}{1.357}     & \multicolumn{1}{l|}{1.388}     & \multicolumn{1}{l|}{-4.7}  \\
\multicolumn{1}{|l|}{1.5} & \multicolumn{1}{l|}{1.8} & \multicolumn{1}{l|}{1.610}     & \multicolumn{1}{l|}{1.623}     & \multicolumn{1}{l|}{-1.4}   & \multicolumn{1}{l|}{1.307}     & \multicolumn{1}{l|}{1.347}     & \multicolumn{1}{l|}{-8.5}  \\
\multicolumn{1}{|l|}{1.8} & \multicolumn{1}{l|}{2.1} & \multicolumn{1}{l|}{1.538}     & \multicolumn{1}{l|}{1.535}     & \multicolumn{1}{l|}{0.5}    & \multicolumn{1}{l|}{1.266}     & \multicolumn{1}{l|}{1.313}     & \multicolumn{1}{l|}{-14.0} \\
\multicolumn{1}{|l|}{2.1} & \multicolumn{1}{l|}{2.4} & \multicolumn{1}{l|}{1.464}     & \multicolumn{1}{l|}{1.514}     & \multicolumn{1}{l|}{-7.4}   & \multicolumn{1}{l|}{1.253}     & \multicolumn{1}{l|}{1.303}     & \multicolumn{1}{l|}{-15.2} \\
\multicolumn{1}{|l|}{2.4} & \multicolumn{1}{l|}{2.7} & \multicolumn{1}{l|}{1.454}     & \multicolumn{1}{l|}{1.473}     & \multicolumn{1}{l|}{-3.5}   & \multicolumn{1}{l|}{1.237}     & \multicolumn{1}{l|}{1.284}     & \multicolumn{1}{l|}{-16.6} \\
\multicolumn{1}{|l|}{2.7} & \multicolumn{1}{l|}{3.0} & \multicolumn{1}{l|}{1.402}     & \multicolumn{1}{l|}{1.457}     & \multicolumn{1}{l|}{-9.5}   & \multicolumn{1}{l|}{1.232}     & \multicolumn{1}{l|}{1.274}     & \multicolumn{1}{l|}{-14.8} \\ \hline
\end{tabular}
\caption{This table summarises the detour for commuting trips and for typical network values by distance ranges between origin and destination. It shows how they significantly differentiate, especially as crow fly distance grows (column $d$). In this table, we look at granular intervals, within the distance band 0-3km (darkred distributions from figures \ref{detour_band_osm} and \ref{detour_band_os}), the full table for up to 15km is given in the appendix.}
\label{detour_sign_3km}
\end{table}

The distribution of all $\delta_{ij,k}$ is shown in figure \ref{detour_distr}. It is not surprising to see that the resulting $\boldsymbol{dodgr}$ networks with a custom weighting profile, which tends to increase the actual distances, show greater values. The results for the remaining networks are more similar, with slightly increasing mean and variance from the whole OSM network, followed by the OSMnx walk and cycle profiles and finally OS. To our knowledge, there is no research linking typical detour values to the density of the nodes and edges of the network, however, it seems intuitively natural for a detour to increase as a consequence of removing edges from a network. Hence, sparser networks are expected to have greater typical detour values. 

We further develop the detour index to account for the flows observed across a set of origin-destination pairs. Let $w_{ij}$ be a flow observed between origin $i$ and destination $j$, by weighting the detour values with the flow, we obtain a more accurate representation of the detours that commuters take on average during their commute: 

\begin{equation} \label{equ:weighted_detour}
    \delta_c = \frac{1}{W}\sum_{i,j \in S,i\neq j} w_{ij} \delta_{ij,k}
\end{equation}

where $W = \sum_{i,j \in S, i\neq j}w_{ij}$ is the total flow between different origins and destinations. 

One of the main characteristics of flows is that their magnitude decreases as we get further from an origin. They also show a wider range of values for shorter distances, spanning several orders of magnitude. At the same time, we observe a decreasing average detour index as the Euclidean distance increases (fig. \ref{detour_dist_band}) between origin and destination, while the potential number of reachable destinations from an origin grows approximately as the square of the distance. These different trends introduce some non trivial relations between flows, their magnitude, the distances on which they occur and their detour values. To further uncover some patterns relating these, we look at the distribution of the mean weighted detour (equation \ref{equ:weighted_detour}) of a trip and the mean detour of the network in 300 metres intervals of Euclidean distance between 0 and 15 km. In this way we minimise the effects of a varying Euclidean distance, and focus only on the typical values of the detour weighted and non weighted by flow. Within each resulting interval we test for the  significance in the difference between the means, and find that it becomes strong (p-value $\approx 0$) inside the interval 900-1200 metres (see table \ref{detour_sign_3km}) and remains statistically significant for all the distances beyond that. Moreover, the t-statistic gets stronger as the Euclidean distance increases. This indicates that for shorter trips, detour has a less significant or no impact on the trip, but as the distance increases, we observe a significant shift of the average weighted detour towards smaller values. This observation can be useful when designing road infrastructure, where aiming to minimise a weighted detour index of a population can potentially have a positive effect on the number of trips done between locations, especially as the Euclidean distance separating points gets greater. 
Figure \ref{detour_dist_band} shows an example distribution of detours with wider intervals for better visualisation purposes. We observe that as the Euclidean distances increase, the distribution of detours narrows down and shifts towards a smaller mean. This behaviour seems to be a fundamental property of the detour index and has been observed for other real world road networks. The full results for granular distance intervals and statistical significance is included in the appendix table \ref{detour_sign_tab}.


\section{Modelling}

This section considers the previous results and integrates the distance matrices produced from the different network profiles and origin-destination centroids into a spatial interaction model. With the available flows data, a doubly constrained gravity model is built and tested.

\subsection{Spatial interaction models}
This family of models, also referred to as gravity models, has been around for a while and was formalised in the late 1960s by \cite{wilson_family_1971}. It allows to build good predictions of flows based on input variables that are relatively easy to find. The name originates from the similarity with the gravity law derived by Isaac Newton. This problem has been reformulated and formalised with statistical considerations of the optimal assignment from a set of origins in a grid to a set of available destinations under a cost value that depends on the distance. 
Intuitively, we consider that there are factors contributing to a greater flow of people between 2 given places and others reducing it. 
Some parameters will thus be directly proportional to the flow of people while others will be inversely so. The factors contributing to the flow positively are usually taken to be the population at the origin location and the job availability at the destination, while the distance between the two locations has a inverse effect on the flow. Formally, this means that $T_{ij} \sim O_iD_jf(d_{ij})$ where $O_i$ is the origin population in location $i$, $D_j$ is the employment at the destination location $j$, $f(d_{ij})$ is a decreasing function of the distance, usually referred to as generalised cost.
The known constraints of the system impose that all the flows originating at a location $i$ must be equal to the local working population, while all the flows arriving to a destination $j$ be equal to the local number of jobs, thus:

\begin{equation} \label{constraints}
    \begin{split}
    O_i = \sum_j T_{ij}, \\
    D_j = \sum_i T_{ij}
    \end{split}
\end{equation}
The travel cost function can be derived \parencite{wilson_family_1971} and has the form $f(d_{ij}) = e^{-\beta d_{ij}}$, where $\beta$ is an exponent that depends on the system and is usually calculated for different modes of transport and geographies. 
Additionally, the origin and destination weighting parameters are obtained through the Lagrange multipliers method, resulting in a equation of the form: 
\begin{equation} \label{doubconstr}
    T_{ij}=A_i O_i B_j D_j e^{-\beta d_{ij}}
\end{equation}

which is referred to as the doubly constrained one since the available information about the system allows to set constraints on both the origin and destination locations. Alternatively, if some information is missing, it is possible to use only one constraint. 

The parameters $A_i,B_j$ are derived from the expression:
\begin{equation}
\begin{split}
     A_i= \lbrack \sum_j B_j D_j e^{-\beta d_{ij}}\rbrack ^{-1} \\
    B_j =  \lbrack \sum_i A_i O_i e^{-\beta d_{ij}}\rbrack ^{-1}
\end{split}
\end{equation}
 with values obtained through an iterative process \parencite{deming_least_1940}.

The flows data is obtained via the census table with origin destination pairs and number of commuters by mode of transport. The distance measures can be obtained in different manners as discussed in section \ref{routing}. The distance matrices obtained are used in the gravity model further on. 

\subsection{Case Study: active travel commuters in London.}
The methods described in the first two sections will now be combined with the formalism of spatial interaction models to have a predictive model of active travel commuter flows in London based on the different networks and centroid types that were introduced. 

\begin{figure}[!h]
\begin{subfigure}[b]{.7\linewidth}
    \includegraphics[width=\linewidth]{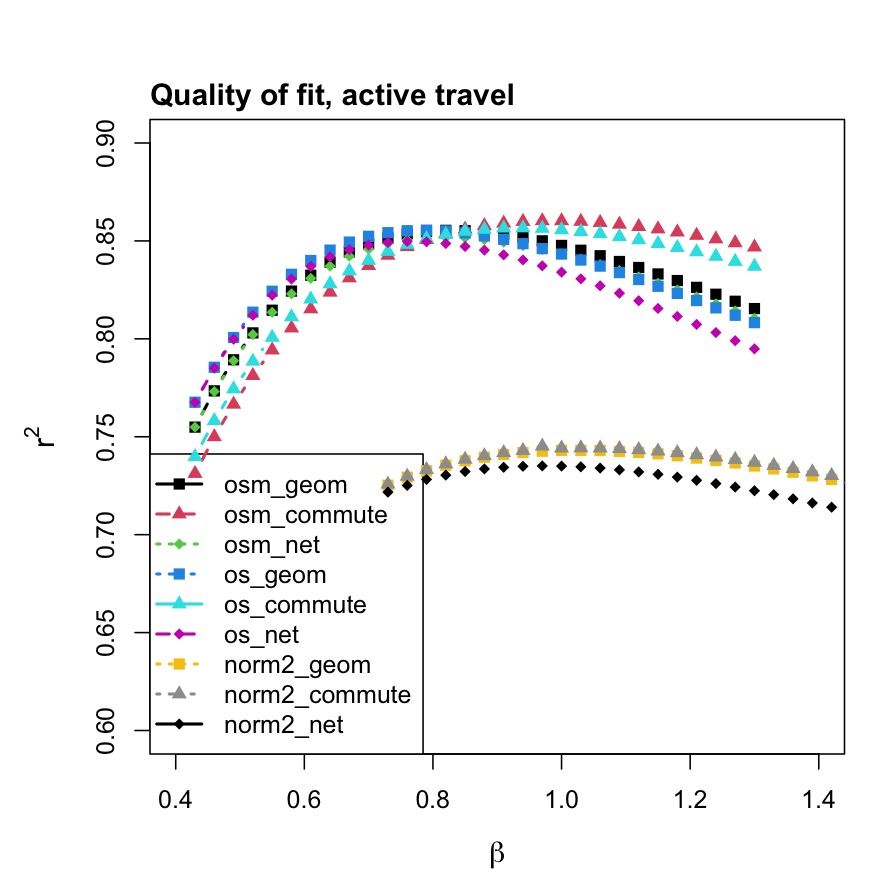}
    \setcounter{subfigure}{0}%
    \caption{}
    \label{quality_of_fit_at}
\end{subfigure}

\begin{subfigure}[b]{.45\linewidth}
    \includegraphics[width=\linewidth]{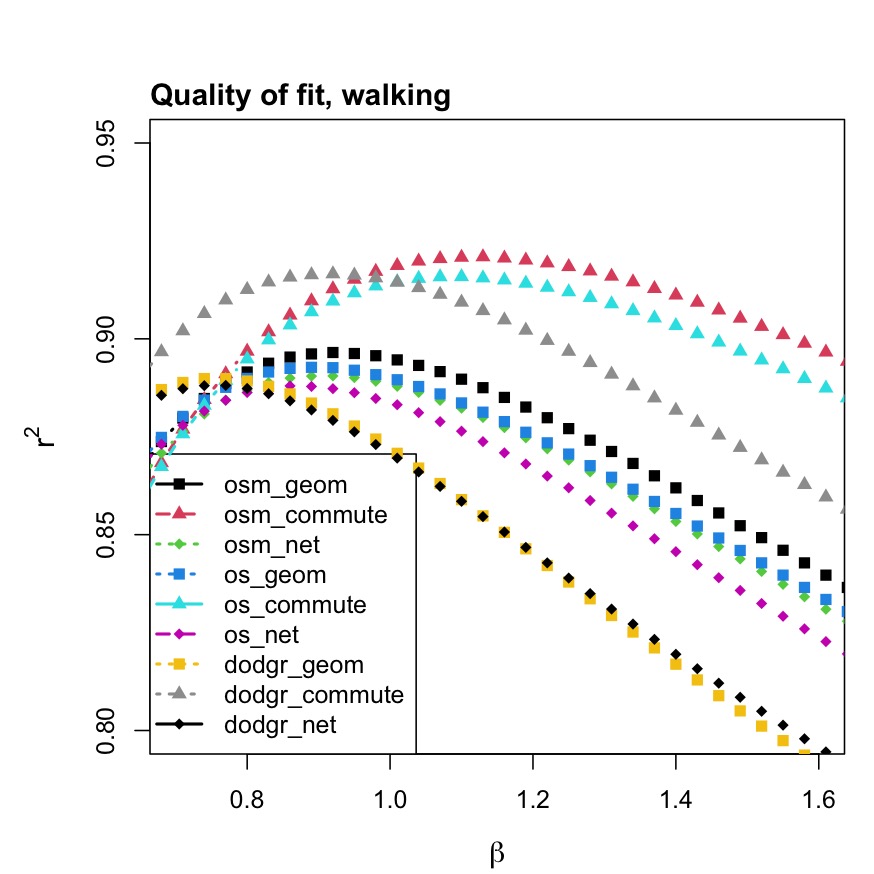}
    \setcounter{subfigure}{1}%
    \caption{}
    \label{qual_fit_walk}
\end{subfigure}
\begin{subfigure}[b]{.45\linewidth}
    \includegraphics[width=\linewidth]{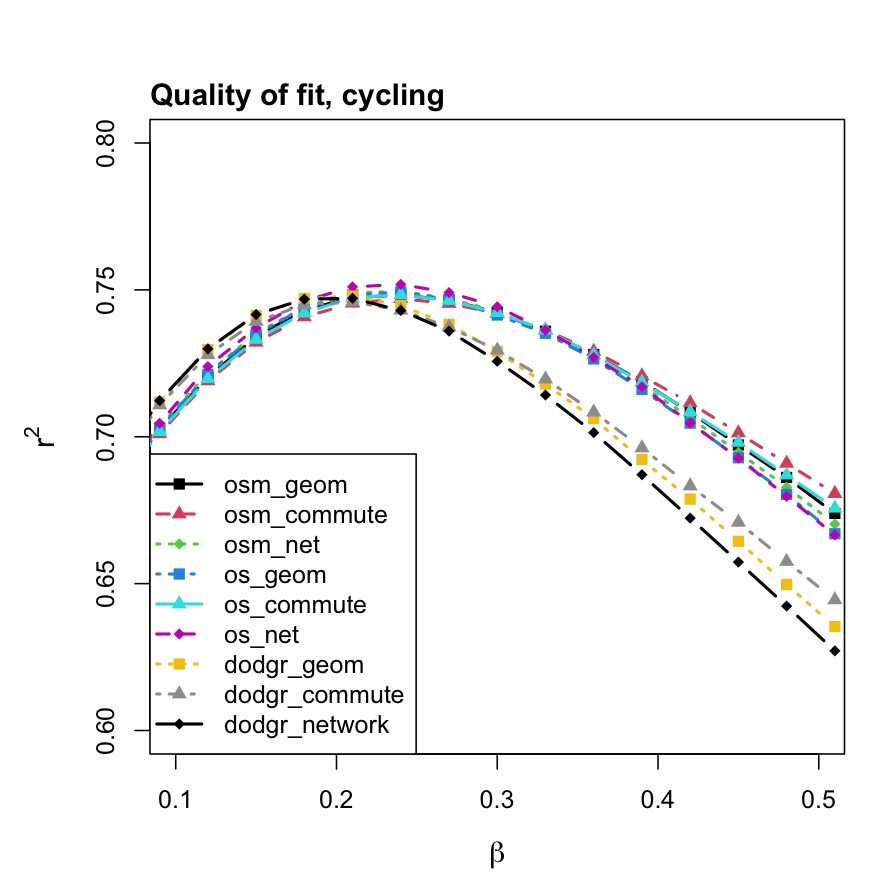}
    \caption{}
    \label{qual_fit_cycle}
\end{subfigure}
\caption{Quality of fit calibration for different networks, mode, centroids. The most significant impact on the quality of fit occurs when changing centroids on the pedestrian model. Each legend element has first mentioned the network from which the cost matrix (\textit{os},\textit{osm},\textit{dodgr},\textit{norm2}), with \textit{norm2} indicating the crow-fly distances matrix. And as second the type of centroids used. }
\label{qual_fit}
\end{figure}

The methods and data described earlier allows for a generalised approach to modelling flows, and when applied to active travel commuters, provide some high quality of fit results in the case study of London. Figure \ref{quality_of_fit_at} shows the results of the calibration of the model to maximise the $r^2$ value. The quality of fit of the models using Euclidean distance between locations as cost is also computed and, despite being lower than network distance, provides a good estimate of the flow with values reaching $r^2=0.75$. One can see that depending on the network, the maximum is reached for differing values, and certain types of networks provide a better stability of the quality of fit under a varying $\beta$ exponent than others. While all the networks reach a high and very similar value of around $r^2 = 0.85$, they do it for different values and the best quality of fit is reached with the OSM and OS networks using \textit{commute} centroids, indicating the importance of the discussion in section \ref{what_to_what} on the knowledge of where the routing is being done. The network and geometric centroids appear less adapted for routing based on the quality of fit obtained for all types of network considered. 

\section{Discussion and conclusion}


In this work, we first conducted a methodological analysis covering network, routing and modelling steps. By comparing different walking and cycling networks and prioritising distance as a critical factor in commuting route choice for active travel, we explored their capacity to estimate real-world flows using a spatial interaction model. We use London as case study, using the 2011 census flows at MSOA level. Throughout the process, we observe that the use of different data sources, network construction, and routing approaches can lead to diverse outcomes in the analysis, both at a broad city-wide scale and at a more local level in different areas of the city. We illustrate the limitations of the OSM data when attempting to select network profiles for specific modes of transport. To counter these limitations, we keep a network that is filtered only to exclude the types of roads that are illegal to use as an active traveller, those are motorway and motorway links and propose a combined active-travel framework allowing to model cycling and walking trips. The results of the spatial interaction model show that this type of network performs best when associated to what we defined as \textit{commute} centroids, accounting for the spatial distribution of residents and jobs inside the census areas. The spatial interaction model that we calibrated on this combination of network and centroid was the best fitting and most stable across the cost function exponent values, highlighting the importance not only of the parameters related to the network itself, but also to the routing process. The quality of fit values for all networks were relatively high, and the best predictions were obtained using the unfiltered OSM and OS networks with commute centroids. 

We further investigate the commuting trip patterns by using the detour index. It shows to be a useful measure for assessing the performance of different networks at minimising the trip length for a given Euclidean distance. We introduce the weighted detour index and show the tendency of London's active travellers to commute in a way that minimises the detour of their trip when the Euclidean distance to their job is beyond 900-1200 metres. Minimising the detour on the way to work seems to be important to active travel commuting, especially as the Euclidean distance between origin and destination increases. We also observe considerable spatial variation in the potential walking and commuting time between MSOAs, which is determined by the uneven distribution and connectivity of the network, as well as the inconsistencies in data sources, which can impact the models that are built on top of them. These findings emphasise the importance of careful consideration and validation of the input data coming from any kind of sources, whether official or crowd sourced. The network with the lowest average detour index also showed the best fitting in the spatial interaction model. This indicates the  possibility to use this index as a guide when designing new infrastructure that aims to connect a population to an employment distribution in a way that minimises the weighted detour of the trips. The main focus of this work is done on active travel as a mean to commute, but the applications and possibilities of the models and methods go beyond that and provide a general framework on open source data for active-travel modelling. Our research has implications for urban planning and policy-making, as it highlights the need for accurate network data and appropriate modelling techniques to inform the development of sustainable transport infrastructure and promote active travel in cities. 

\section{Acknowledgements}

This paper has received funding from the QUANT and RUBICON projects from the Alan Turing Institute under grant
TU/ASG/R-SPEU-102.

\printbibliography

@Manual{microbench,
    title = {microbenchmark: Accurate Timing Functions},
    author = {Olaf Mersmann},
    year = {2023},
    note = {R package version 1.4.10},
    url = {https://CRAN.R-project.org/package=microbenchmark},
  }

@article{Cervero2019,
author = {Cervero, Robert and Denman, Steve and Jin, Ying},
doi = {10.1016/j.tranpol.2018.09.007},
issn = {1879310X},
journal = {Transport Policy},
number = {September 2018},
pages = {153--164},
publisher = {Elsevier Ltd},
title = {{Network design, built and natural environments, and bicycle commuting: Evidence from British cities and towns}},
url = {https://doi.org/10.1016/j.tranpol.2018.09.007},
volume = {74},
year = {2019}
}

@article{Schoner2014,
author = {Schoner, Jessica and Cao, Xinyu},
doi = {10.3141/2464-09},
isbn = {9780309295567},
issn = {03611981},
journal = {Transportation Research Record},
pages = {67--76},
title = {{Walking for purpose and pleasure influences of light rail, built environment, and residential self-selection on pedestrian travel}},
volume = {2464},
year = {2014}
}

@article{Broach2011,
author = {Broach, Joseph and Gliebe, John and Dill, Jennifer},
isbn = {5037258480},
journal = {TRB 2011 Annual Meeting},
pmid = {1329018},
title = {{Bicycle route choice model developed using revealed preference GPS data}},
url = {ftp://ftp.hsrc.unc.edu/pub/TRB2011/data/papers/11-3901.pdf},
volume = {5464},
year = {2011}
}

@article{Winters2010,
author = {Winters, Meghan and Teschke, Kay and Grant, Michael and Setton, Eleanor M. and Brauer, Michael},
doi = {10.3141/2190-01},
issn = {03611981},
journal = {Transportation Research Record},
number = {2190},
pages = {1--10},
title = {{How far out of the way will we travel? Built environment influences on route selection for bicycle and car travel}},
year = {2010}
}

@article{Krizek2009,
author = {Krizek, Kevin J. and Handy, Susan L. and Forsyth, Ann},
doi = {10.1068/b34023},
issn = {14723417},
journal = {Environment and Planning B: Planning and Design},
number = {4},
pages = {725--740},
title = {{Explaining changes in walking and bicycling behavior: Challenges for transportation research}},
volume = {36},
year = {2009}
}

@Misc{cpprouting,
    title = {{cppRouting}: Fast Implementation of Dijkstra Algorithm in R},
    author = {Vincent Larmet},
    year = {2019},
    url = {https://github.com/vlarmet/cppRouting},
  }

@Article{dodgr,
    journal = {Transport Findings},
    doi = {10.32866/6945},
    publisher = {Network Design Lab},
    title = {dodgr: An R package for network flow aggregation},
    author = {Mark Padgham},
    year = {2019},
    month = {2},
  }

@article{wilson_family_1971,
	title = {A {Family} of {Spatial} {Interaction} {Models}, and {Associated} {Developments}},
	volume = {3},
	issn = {0308-518X, 1472-3409},
	url = {http://journals.sagepub.com/doi/10.1068/a030001},
	doi = {10.1068/a030001},
	language = {en},
	number = {1},
	urldate = {2021-12-26},
	journal = {Environment and Planning A: Economy and Space},
	author = {Wilson, A G},
	month = mar,
	year = {1971},
	pages = {1--32},
}

@article{deming_least_1940,
	title = {On a {Least} {Squares} {Adjustment} of a {Sampled} {Frequency} {Table} {When} the {Expected} {Marginal} {Totals} are {Known}},
	volume = {11},
	issn = {0003-4851},
	url = {http://projecteuclid.org/euclid.aoms/1177731829},
	doi = {10.1214/aoms/1177731829},
	language = {en},
	number = {4},
	urldate = {2022-05-25},
	journal = {The Annals of Mathematical Statistics},
	author = {Deming, W. Edwards and Stephan, Frederick F.},
	month = dec,
	year = {1940},
	pages = {427--444},
}

@book{stein_introduction_2009,
	series = {The {MIT} {Press}},
	title = {Introduction to algorithms},
	isbn = {978-0-262-03384-8},
	publisher = {The MIT Press},
	author = {Stein, Clifford and Leiserson, Charles E and Cormen, Thomas H and Rivest, Ronald L},
	year = {2009},
}

@article{boeing_OSMnx_2017,
	title = {{OSMnx}: {New} methods for acquiring, constructing, analyzing, and visualizing complex street networks},
	volume = {65},
	issn = {01989715},
	shorttitle = {{OSMnx}},
	url = {https://linkinghub.elsevier.com/retrieve/pii/S0198971516303970},
	doi = {10.1016/j.compenvurbsys.2017.05.004},
	language = {en},
	urldate = {2022-05-19},
	journal = {Computers, Environment and Urban Systems},
	author = {Boeing, Geoff},
	month = sep,
	year = {2017},
	pages = {126--139},
}

@Manual{sfnetworks,
    title = {sfnetworks: Tidy Geospatial Networks},
    author = {Lucas {van der Meer} and Lorena Abad and Andrea Gilardi and Robin Lovelace},
    year = {2022},
    note = {R package version 0.5.5},
    url = {https://CRAN.R-project.org/package=sfnetworks},
  }

@Manual{tidygraph,
    title = {tidygraph: A Tidy API for Graph Manipulation},
    author = {Thomas Lin Pedersen},
    year = {2022},
    note = {R package version 1.2.1},
    url = {https://CRAN.R-project.org/package=tidygraph},
  }

@Article{igraph,
    title = {The igraph software package for complex network research},
    author = {Gabor Csardi and Tamas Nepusz},
    journal = {InterJournal},
    volume = {Complex Systems},
    pages = {1695},
    year = {2006},
    url = {https://igraph.org},
  }

@article{strano_elementary_2012,
	title = {Elementary processes governing the evolution of road networks},
	volume = {2},
	issn = {2045-2322},
	url = {http://www.nature.com/articles/srep00296},
	doi = {10.1038/srep00296},
	pages = {296},
	number = {1},
	journaltitle = {Scientific Reports},
	shortjournal = {Sci Rep},
	author = {Strano, Emanuele and Nicosia, Vincenzo and Latora, Vito and Porta, Sergio and Barthélemy, Marc},
	urldate = {2021-04-05},
	date = {2012-12},
	langid = {english},
	note = {Number: 1},
}

@article{masucci_exploring_2014,
	title = {Exploring the evolution of London's street network in the information space: A dual approach},
	volume = {89},
	issn = {1539-3755, 1550-2376},
	url = {https://link.aps.org/doi/10.1103/PhysRevE.89.012805},
	doi = {10.1103/PhysRevE.89.012805},
	shorttitle = {Exploring the evolution of London's street network in the information space},
	pages = {012805},
	number = {1},
	journaltitle = {Physical Review E},
	shortjournal = {Phys. Rev. E},
	author = {Masucci, A. Paolo and Stanilov, Kiril and Batty, Michael},
	urldate = {2021-06-26},
	date = {2014-01-13},
	langid = {english},
	note = {Number: 1},
}

@Article{osmdata,
  title = {osmdata},
  author = {Mark Padgham and Bob Rudis and Robin Lovelace and Maëlle Salmon},
  journal = {The Journal of Open Source Software},
  year = {2017},
  volume = {2},
  number = {14},
  month = {jun},
  publisher = {The Open Journal},
  url = {https://doi.org/10.21105/joss.00305},
  doi = {10.21105/joss.00305},
}

@Manual{osmextract,
  title = {osmextract: Download and Import Open Street Map Data Extracts},
  author = {Andrea Gilardi and Robin Lovelace},
  year = {2022},
  note = {https://docs.ropensci.org/osmextract/, https://github.com/ropensci/osmextract},
}

@article{hidalgo_trillion_2020,
	title = {Trillion dollar streets},
	volume = {47},
	issn = {2399-8083, 2399-8091},
	url = {http://journals.sagepub.com/doi/10.1177/2399808320949295},
	doi = {10.1177/2399808320949295},
	pages = {1133--1135},
	number = {7},
	journaltitle = {Environment and Planning B: Urban Analytics and City Science},
	shortjournal = {Environment and Planning B: Urban Analytics and City Science},
	author = {Hidalgo, César A},
	urldate = {2022-12-05},
	date = {2020-09},
	langid = {english},
}

@article{fleury_geospatial_2021,
	title = {Geospatial analysis of individual-based Parkinson's disease data supports a link with air pollution: A case-control study},
	volume = {83},
	issn = {13538020},
	url = {https://linkinghub.elsevier.com/retrieve/pii/S135380202100002X},
	doi = {10.1016/j.parkreldis.2020.12.013},
	shorttitle = {Geospatial analysis of individual-based Parkinson's disease data supports a link with air pollution},
	pages = {41--48},
	journaltitle = {Parkinsonism \& Related Disorders},
	shortjournal = {Parkinsonism \& Related Disorders},
	author = {Fleury, Vanessa and Himsl, Rebecca and Joost, Stéphane and Nicastro, Nicolas and Bereau, Matthieu and Guessous, Idris and Burkhard, Pierre R.},
	urldate = {2021-05-14},
	date = {2021-02},
	langid = {english},
}

@article{boeing_street_2021,
	title = {Street Network Models and Indicators for Every Urban Area in the World},
	issn = {0016-7363, 1538-4632},
	url = {https://onlinelibrary.wiley.com/doi/10.1111/gean.12281},
	doi = {10.1111/gean.12281},
	pages = {gean.12281},
	journaltitle = {Geographical Analysis},
	shortjournal = {Geogr Anal},
	author = {Boeing, Geoff},
	urldate = {2022-02-10},
	date = {2021-03-09},
	langid = {english},
}

@article{liu_generalized_2021,
	title = {A Generalized Framework for Measuring Pedestrian Accessibility around the World Using Open Data},
	issn = {0016-7363, 1538-4632},
	url = {https://onlinelibrary.wiley.com/doi/10.1111/gean.12290},
	doi = {10.1111/gean.12290},
	pages = {gean.12290},
	journaltitle = {Geographical Analysis},
	shortjournal = {Geogr Anal},
	author = {Liu, Shiqin and Higgs, Carl and Arundel, Jonathan and Boeing, Geoff and Cerdera, Nicholas and Moctezuma, David and Cerin, Ester and Adlakha, Deepti and Lowe, Melanie and Giles‐Corti, Billie},
	urldate = {2022-05-25},
	date = {2021-05-19},
	langid = {english},
}

@article{Kelly2014,
author = {Kelly, Paul and Kahlmeier, Sonja and G{\"{o}}tschi, Thomas and Orsini, Nicola and Richards, Justin and Roberts, Nia and Scarborough, Peter and Foster, Charlie},
doi = {10.1186/s12966-014-0132-x},
issn = {14795868},
journal = {International Journal of Behavioral Nutrition and Physical Activity},
number = {1},
pmid = {25344355},
title = {{Systematic review and meta-analysis of reduction in all-cause mortality from walking and cycling and shape of dose response relationship}},
volume = {11},
year = {2014}
}

@article{Oja2011,
author = {Oja, P. and Titze, S. and Bauman, A. and de Geus, B. and Krenn, P. and Reger-Nash, B. and Kohlberger, T.},
doi = {10.1111/j.1600-0838.2011.01299.x},
issn = {09057188},
journal = {Scandinavian Journal of Medicine and Science in Sports},
number = {4},
pages = {496--509},
pmid = {21496106},
title = {{Health benefits of cycling: A systematic review}},
volume = {21},
year = {2011}
}

@article{Saunders2013,
author = {Saunders, Lucinda E. and Green, Judith M. and Petticrew, Mark P. and Steinbach, Rebecca and Roberts, Helen},
doi = {10.1371/journal.pone.0069912},
issn = {19326203},
journal = {PLoS ONE},
number = {8},
pmid = {23967064},
title = {{What Are the Health Benefits of Active Travel? A Systematic Review of Trials and Cohort Studies}},
volume = {8},
year = {2013}
}

@article{porta2009street,
  title={Street centrality and densities of retail and services in Bologna, Italy},
  author={Porta, Sergio and Strano, Emanuele and Iacoviello, Valentino and Messora, Roberto and Latora, Vito and Cardillo, Alessio and Wang, Fahui and Scellato, Salvatore},
  journal={Environment and Planning B: Planning and design},
  volume={36},
  number={3},
  pages={450--465},
  year={2009},
  publisher={SAGE Publications Sage UK: London, England}
}

@article{porta2012street,
  title={Street centrality and the location of economic activities in Barcelona},
  author={Porta, Sergio and Latora, Vito and Wang, Fahui and Rueda, Salvador and Strano, Emanuele and Scellato, Salvatore and Cardillo, Alessio and Belli, Eugenio and Cardenas, Francisco and Cormenzana, Berta and others},
  journal={Urban Studies},
  volume={49},
  number={7},
  pages={1471--1488},
  year={2012},
  publisher={SAGE Publications Sage UK: London, England}
}

@inproceedings{law2013measuring,
  title={Measuring the influence of spatial configuration on the housing market in metropolitan London},
  author={Law, Stephen and Karimi, Kayvan and Penn, Alan and Chiaradia, Alain},
  booktitle={Proceedings of the 2013 International Space Syntax Symposium, Seoul, Korea},
  volume={31},
  year={2013}
}

@article{Piovani2017,
archivePrefix = {arXiv},
arxivId = {1703.10419},
author = {Piovani, Duccio and Molinero, Carlos and Wilson, Alan},
doi = {10.1371/journal.pone.0185787},
eprint = {1703.10419},
isbn = {1111111111},
issn = {19326203},
journal = {PLoS ONE},
number = {10},
title = {{Urban retail location: Insights from percolation theory and spatial interaction modeling}},
volume = {12},
year = {2017}
}

@article{hillier2012normalising,
  title={Normalising least angle choice in Depthmap-and how it opens up new perspectives on the global and local analysis of city space},
  author={Hillier, Bill and Yang, Tao and Turner, Alasdair},
  journal={Journal of Space syntax},
  volume={3},
  number={2},
  pages={155--193},
  year={2012},
  publisher={University College London}
}

@incollection{vaughan2007spatial,
  title={The spatial form of poverty in Charles Booth's London},
  author={Vaughan, Laura},
  year={2007},
  publisher={Elsevier}
}

@article{molinero2015fractured,
  title={The fractured nature of British politics},
  author={Molinero, Carlos and Arcaute, Elsa and Smith, Duncan and Batty, Michael},
  journal={arXiv preprint arXiv:1505.00217},
  year={2015}
}

@article{freiria2015multiscale,
  title={The multiscale importance of road segments in a network disruption scenario: A risk-based approach},
  author={Freiria, Susana and Tavares, Alexandre O and Pedro Juli{\~a}o, Rui},
  journal={Risk analysis},
  volume={35},
  number={3},
  pages={484--500},
  year={2015},
  publisher={Wiley Online Library}
}

@article{novak2014link,
  title={A link-focused methodology for evaluating accessibility to emergency services},
  author={Novak, David C and Sullivan, James L},
  journal={Decision Support Systems},
  volume={57},
  pages={309--319},
  year={2014},
  publisher={Elsevier}
}

@article{porta2006network,
  title={The network analysis of urban streets: a primal approach},
  author={Porta, Sergio and Crucitti, Paolo and Latora, Vito},
  journal={Environment and Planning B: planning and design},
  volume={33},
  number={5},
  pages={705--725},
  year={2006},
  publisher={SAGE Publications Sage UK: London, England}
}

@article{strano2013urban,
  title={Urban street networks, a comparative analysis of ten European cities},
  author={Strano, Emanuele and Viana, Matheus and da Fontoura Costa, Luciano and Cardillo, Alessio and Porta, Sergio and Latora, Vito},
  journal={Environment and Planning B: Planning and Design},
  volume={40},
  number={6},
  pages={1071--1086},
  year={2013},
  publisher={SAGE Publications Sage UK: London, England}
}

@article{marshall2018street,
  title={Street network studies: from networks to models and their representations},
  author={Marshall, Stephen and Gil, Jorge and Kropf, Karl and Tomko, Martin and Figueiredo, Lucas},
  journal={Networks and Spatial Economics},
  pages={1--15},
  year={2018},
  publisher={Springer}
}

@article{Rhoads2020,
archivePrefix = {arXiv},
arxivId = {2009.12548},
author = {Rhoads, Daniel and Sol{\'{e}}-Ribalta, Albert and Gonz{\'{a}}lez, Marta C. and Borge-Holthoefer, Javier},
eprint = {2009.12548},
title = {{Planning for sustainable Open Streets in pandemic cities}},
url = {http://arxiv.org/abs/2009.12548},
year = {2020}
}

@article{Mahfouz2021,
archivePrefix = {arXiv},
arxivId = {2105.03712},
author = {Mahfouz, Hussein and Arcaute, Elsa and Lovelace, Robin},
eprint = {2105.03712},
pages = {1--31},
title = {{A Road Segment Prioritization Approach for Cycling Infrastructure}},
url = {http://arxiv.org/abs/2105.03712},
year = {2021}
}

@article{Penn2003,
author = {Penn, Alan},
doi = {10.1177/0013916502238864},
issn = {00139165},
journal = {Environment and Behavior},
number = {1},
pages = {30--65},
title = {{Space syntax and spatial cognition: Or why the axial line?}},
volume = {35},
year = {2003}
}

@article{bassolas_hierarchical_2019,
	title = {Hierarchical organization of urban mobility and its connection with city livability},
	volume = {10},
	issn = {2041-1723},
	url = {http://www.nature.com/articles/s41467-019-12809-y},
	doi = {10.1038/s41467-019-12809-y},
	number = {1},
	urldate = {2021-02-25},
	journal = {Nature Communications},
	author = {Bassolas, Aleix and Barbosa-Filho, Hugo and Dickinson, Brian and Dotiwalla, Xerxes and Eastham, Paul and Gallotti, Riccardo and Ghoshal, Gourab and Gipson, Bryant and Hazarie, Surendra A. and Kautz, Henry and Kucuktunc, Onur and Lieber, Allison and Sadilek, Adam and Ramasco, José J.},
	month = dec,
	year = {2019},
	pages = {4817},
}

@article{gallotti_unraveling_2021,
	title = {Unraveling the hidden organisation of urban systems and their mobility flows},
	volume = {10},
	issn = {2193-1127},
	url = {https://epjdatascience.springeropen.com/articles/10.1140/epjds/s13688-020-00258-3},
	doi = {10.1140/epjds/s13688-020-00258-3},
	number = {1},
	urldate = {2021-03-01},
	journal = {EPJ Data Science},
	author = {Gallotti, Riccardo and Bertagnolli, Giulia and De Domenico, Manlio},
	month = dec,
	year = {2021},
	pages = {3},
}

@article{barthelemy_paths_2017,
	title = {From paths to blocks: {New} measures for street patterns},
	volume = {44},
	issn = {2399-8083, 2399-8091},
	shorttitle = {From paths to blocks},
	url = {http://journals.sagepub.com/doi/10.1177/0265813515599982},
	doi = {10.1177/0265813515599982},
	number = {2},
	urldate = {2021-07-29},
	journal = {Environment and Planning B: Urban Analytics and City Science},
	author = {Barthelemy, Marc},
	month = mar,
	year = {2017},
	pages = {256--271},
}

@article{zhong_detecting_2014,
	title = {Detecting the dynamics of urban structure through spatial network analysis},
	volume = {28},
	issn = {1365-8816, 1362-3087},
	url = {http://www.tandfonline.com/doi/full/10.1080/13658816.2014.914521},
	doi = {10.1080/13658816.2014.914521},
	number = {11},
	urldate = {2021-09-05},
	journal = {International Journal of Geographical Information Science},
	author = {Zhong, Chen and Arisona, Stefan Müller and Huang, Xianfeng and Batty, Michael and Schmitt, Gerhard},
	month = nov,
	year = {2014},
	pages = {2178--2199},
}

@article{barthelemy_spatial_2011,
	title = {Spatial networks},
	volume = {499},
	issn = {03701573},
	url = {https://linkinghub.elsevier.com/retrieve/pii/S037015731000308X},
	doi = {10.1016/j.physrep.2010.11.002},
	number = {1},
	urldate = {2021-07-02},
	journal = {Physics Reports},
	author = {Barthélemy, Marc},
	month = feb,
	year = {2011},
	pages = {1--101},
}

@article{lovelace_propensity_2017,
	title = {The {Propensity} to {Cycle} {Tool}: {An} open source online system for sustainable transport planning},
	volume = {10},
	issn = {1938-7849},
	shorttitle = {The {Propensity} to {Cycle} {Tool}},
	url = {https://www.jtlu.org/index.php/jtlu/article/view/862},
	doi = {10.5198/jtlu.2016.862},
	number = {1},
	urldate = {2023-03-15},
	journal = {Journal of Transport and Land Use},
	author = {Lovelace, Robin and Goodman, Anna and Aldred, Rachel and Berkoff, Nikolai and Abbas, Ali and Woodcock, James},
	month = jan,
	year = {2017},
}

@article{louf_typology_2014,
	title = {A typology of street patterns},
	volume = {11},
	issn = {1742-5689, 1742-5662},
	url = {https://royalsocietypublishing.org/doi/10.1098/rsif.2014.0924},
	doi = {10.1098/rsif.2014.0924},
	number = {101},
	urldate = {2021-04-05},
	journal = {Journal of The Royal Society Interface},
	author = {Louf, Rémi and Barthelemy, Marc},
	month = dec,
	year = {2014},
	pages = {20140924},
}

@article{palominos_examining_2022,
	title = {Examining the geometry of streets through accessibility: new insights from streetspace allocation analysis},
	issn = {2399-8083, 2399-8091},
	shorttitle = {Examining the geometry of streets through accessibility},
	url = {http://journals.sagepub.com/doi/10.1177/23998083221139849},
	doi = {10.1177/23998083221139849},
	language = {en},
	urldate = {2023-03-15},
	journal = {Environment and Planning B: Urban Analytics and City Science},
	author = {Palominos, Nicolas and Smith, Duncan A},
	month = nov,
	year = {2022},
	pages = {239980832211398},
}

@article{costa_circuity_2021,
	title = {A {Circuity} {Temporal} {Analysis} of {Urban} {Street} {Networks} {Using} {Open} {Data}: {A} {Lisbon} {Case} {Study}},
	volume = {10},
	issn = {2220-9964},
	shorttitle = {A {Circuity} {Temporal} {Analysis} of {Urban} {Street} {Networks} {Using} {Open} {Data}},
	url = {https://www.mdpi.com/2220-9964/10/7/453},
	doi = {10.3390/ijgi10070453},
	language = {en},
	number = {7},
	urldate = {2023-03-15},
	journal = {ISPRS International Journal of Geo-Information},
	author = {Costa, Miguel and Marques, Manuel and Moura, Filipe},
	month = jul,
	year = {2021},
	pages = {453},
}

@article{levinson_minimum_2009,
	title = {The minimum circuity frontier and the journey to work},
	volume = {39},
	issn = {01660462},
	url = {https://linkinghub.elsevier.com/retrieve/pii/S0166046209000660},
	doi = {10.1016/j.regsciurbeco.2009.07.003},
	language = {en},
	number = {6},
	urldate = {2023-03-16},
	journal = {Regional Science and Urban Economics},
	author = {Levinson, David and El-Geneidy, Ahmed},
	month = nov,
	year = {2009},
	pages = {732--738},
}

@article{romanillos_pulse_2018,
	title = {The pulse of the cycling city: visualising {Madrid} bike share system {GPS} routes and cycling flow},
	volume = {14},
	issn = {1744-5647},
	shorttitle = {The pulse of the cycling city},
	url = {https://www.tandfonline.com/doi/full/10.1080/17445647.2018.1438932},
	doi = {10.1080/17445647.2018.1438932},
	language = {en},
	number = {1},
	urldate = {2023-03-13},
	journal = {Journal of Maps},
	author = {Romanillos, Gustavo and Moya-Gómez, Borja and Zaltz-Austwick, Martin and Lamíquiz-Daudén, Patxi J.},
	month = jan,
	year = {2018},
	pages = {34--43},
}

@article{HealthWalk,
    doi = {10.1371/journal.pone.0051462},
    author = {Woodcock, James AND Givoni, Moshe AND Morgan, Andrei Scott},
    journal = {PLOS ONE},
    publisher = {Public Library of Science},
    title = {Health Impact Modelling of Active Travel Visions for England and Wales Using an Integrated Transport and Health Impact Modelling Tool (ITHIM)},
    year = {2013},
    month = {01},
    volume = {8},
    url = {https://doi.org/10.1371/journal.pone.0051462},
    pages = {1-17},
    number = {1},

}

@article {Hamer238,
	author = {Hamer, M and Chida, Y},
	title = {Walking and primary prevention: a meta-analysis of prospective cohort studies},
	volume = {42},
	number = {4},
	pages = {238--243},
	year = {2008},
	doi = {10.1136/bjsm.2007.039974},
	publisher = {British Association of Sport and Excercise Medicine},
	issn = {0306-3674},
	URL = {https://bjsm.bmj.com/content/42/4/238},
	eprint = {https://bjsm.bmj.com/content/42/4/238.full.pdf},
	journal = {British Journal of Sports Medicine}
}

@article{dill_bicycle_2003,
	title = {Bicycle {Commuting} and {Facilities} in {Major} {U}.{S}. {Cities}: {If} {You} {Build} {Them}, {Commuters} {Will} {Use} {Them}},
	volume = {1828},
	issn = {0361-1981, 2169-4052},
	shorttitle = {Bicycle {Commuting} and {Facilities} in {Major} {U}.{S}. {Cities}},
	url = {http://journals.sagepub.com/doi/10.3141/1828-14},
	doi = {10.3141/1828-14},
	language = {en},
	number = {1},
	urldate = {2023-03-23},
	journal = {Transportation Research Record: Journal of the Transportation Research Board},
	author = {Dill, Jennifer and Carr, Theresa},
	month = jan,
	year = {2003},
	pages = {116--123},
}

@article{thompson_sargoni_neighbourhood-level_2023,
	title = {Neighbourhood-level pedestrian navigation using the construal level theory},
	issn = {2399-8083, 2399-8091},
	url = {http://journals.sagepub.com/doi/10.1177/23998083231158371},
	doi = {10.1177/23998083231158371},
	language = {en},
	urldate = {2023-06-08},
	journal = {Environment and Planning B: Urban Analytics and City Science},
	author = {Thompson Sargoni, Obi and Manley, Ed},
	month = feb,
	year = {2023},
	pages = {239980832311583},
}

@misc{office_for_national_statistics_2011_2017,
	title = {2011 {Census} aggegate data ({Data} downloaded: 1 {February} 2017)},
	shorttitle = {2011 {Census} aggegate data ({Data} downloaded},
	url = {https://beta.ukdataservice.ac.uk/datacatalogue/studies/study?id=7427},
	doi = {10.5257/CENSUS/AGGREGATE-2011-2},
	urldate = {2023-06-30},
	publisher = {UK Data Service. Office for National Statistics},
	author = {ONS},
	year = {2017},
}

@techreport{boeing_morphology_2018,
	type = {preprint},
	title = {The {Morphology} and {Circuity} of {Walkable} and {Drivable} {Street} {Networks}},
	url = {https://osf.io/edj2s},
	urldate = {2023-06-13},
	institution = {SocArXiv},
	author = {Boeing, Geoff},
	month = feb,
	year = {2018},
	doi = {10.31235/osf.io/edj2s},
}

@article{yang_universal_2018,
	title = {A universal distribution law of network detour ratios},
	volume = {96},
	issn = {0968090X},
	url = {https://linkinghub.elsevier.com/retrieve/pii/S0968090X18311185},
	doi = {10.1016/j.trc.2018.09.012},
	language = {en},
	urldate = {2023-06-12},
	journal = {Transportation Research Part C: Emerging Technologies},
	author = {Yang, Hai and Ke, Jintao and Ye, Jieping},
	month = nov,
	year = {2018},
	pages = {22--37},
}

@article{jungGravityModelKorean2008,
	title = {Gravity model in the {Korean} highway},
	volume = {81},
	issn = {0295-5075, 1286-4854},
	url = {https://iopscience.iop.org/article/10.1209/0295-5075/81/48005},
	doi = {10.1209/0295-5075/81/48005},
	number = {4},
	urldate = {2023-07-28},
	journal = {EPL (Europhysics Letters)},
	author = {Jung, Woo-Sung and Wang, Fengzhong and Stanley, H. Eugene},
	month = feb,
	year = {2008},
	pages = {48005},
}

\newpage
\section{Appendix}

\subsection{Online material and reproducibility}
The main code and process to access the data are described here. Limitations due to the accessibility of the census flow forbid us to provide it here, but an institutional account allows to connect and download the detailed msoa level flows data from the WICID \footnote{\url{https://wicid.ukdataservice.ac.uk}} portal via an account created on UKDataServices \footnote{\url{https://ukdataservice.ac.uk}}. 

Code: \url{https://github.com/ischlo/quant_cycle_walk}

Additionally, a mini package called \textbf{cppSim} was developed to perform fast and memory efficient spatial interaction models, it is made available in the following github repository:  
\url{https://github.com/ischlo/cppSim}

\newpage

\begin{table}[]
\centering
\begin{tabular}{llllllll}
\multicolumn{2}{c}{\textit{d (km)}}                    & \multicolumn{3}{c}{\textit{OS}}                                                               & \multicolumn{3}{c}{\textit{OSM}}                                                             \\
\multicolumn{1}{c}{min}    & \multicolumn{1}{c|}{max}  & \multicolumn{1}{c}{$\delta_c$} & \multicolumn{1}{c}{$\delta_n$} & \multicolumn{1}{c|}{t-test} & \multicolumn{1}{c}{$\delta_c$} & \multicolumn{1}{c}{$\delta_n$} & \multicolumn{1}{c}{t-test} \\ \hline
\multicolumn{1}{|l|}{0.0}  & \multicolumn{1}{l|}{0.6}  & \multicolumn{1}{l|}{2.113}     & \multicolumn{1}{l|}{2.108}     & \multicolumn{1}{l|}{0.1}    & \multicolumn{1}{l|}{1.498}     & \multicolumn{1}{l|}{1.518}     & \multicolumn{1}{l|}{-0.8}  \\
\multicolumn{1}{|l|}{0.6}  & \multicolumn{1}{l|}{0.9}  & \multicolumn{1}{l|}{1.931}     & \multicolumn{1}{l|}{1.914}     & \multicolumn{1}{l|}{0.9}    & \multicolumn{1}{l|}{1.454}     & \multicolumn{1}{l|}{1.471}     & \multicolumn{1}{l|}{-1.6}  \\
\multicolumn{1}{|l|}{0.9}  & \multicolumn{1}{l|}{1.2}  & \multicolumn{1}{l|}{1.780}     & \multicolumn{1}{l|}{1.791}     & \multicolumn{1}{l|}{-0.7}   & \multicolumn{1}{l|}{1.384}     & \multicolumn{1}{l|}{1.423}     & \multicolumn{1}{l|}{-4.9}  \\
\multicolumn{1}{|l|}{1.2}  & \multicolumn{1}{l|}{1.5}  & \multicolumn{1}{l|}{1.666}     & \multicolumn{1}{l|}{1.700}     & \multicolumn{1}{l|}{-3.0}   & \multicolumn{1}{l|}{1.357}     & \multicolumn{1}{l|}{1.388}     & \multicolumn{1}{l|}{-4.7}  \\
\multicolumn{1}{|l|}{1.5}  & \multicolumn{1}{l|}{1.8}  & \multicolumn{1}{l|}{1.610}     & \multicolumn{1}{l|}{1.623}     & \multicolumn{1}{l|}{-1.4}   & \multicolumn{1}{l|}{1.307}     & \multicolumn{1}{l|}{1.347}     & \multicolumn{1}{l|}{-8.5}  \\
\multicolumn{1}{|l|}{1.8}  & \multicolumn{1}{l|}{2.1}  & \multicolumn{1}{l|}{1.538}     & \multicolumn{1}{l|}{1.535}     & \multicolumn{1}{l|}{0.5}    & \multicolumn{1}{l|}{1.266}     & \multicolumn{1}{l|}{1.313}     & \multicolumn{1}{l|}{-14.0} \\
\multicolumn{1}{|l|}{2.1}  & \multicolumn{1}{l|}{2.4}  & \multicolumn{1}{l|}{1.464}     & \multicolumn{1}{l|}{1.514}     & \multicolumn{1}{l|}{-7.4}   & \multicolumn{1}{l|}{1.253}     & \multicolumn{1}{l|}{1.303}     & \multicolumn{1}{l|}{-15.2} \\
\multicolumn{1}{|l|}{2.4}  & \multicolumn{1}{l|}{2.7}  & \multicolumn{1}{l|}{1.454}     & \multicolumn{1}{l|}{1.473}     & \multicolumn{1}{l|}{-3.5}   & \multicolumn{1}{l|}{1.237}     & \multicolumn{1}{l|}{1.284}     & \multicolumn{1}{l|}{-16.6} \\
\multicolumn{1}{|l|}{2.7}  & \multicolumn{1}{l|}{3.0}  & \multicolumn{1}{l|}{1.402}     & \multicolumn{1}{l|}{1.457}     & \multicolumn{1}{l|}{-9.5}   & \multicolumn{1}{l|}{1.232}     & \multicolumn{1}{l|}{1.274}     & \multicolumn{1}{l|}{-14.8} \\
\multicolumn{1}{|l|}{3.0}  & \multicolumn{1}{l|}{3.3}  & \multicolumn{1}{l|}{1.377}     & \multicolumn{1}{l|}{1.430}     & \multicolumn{1}{l|}{-10.8}  & \multicolumn{1}{l|}{1.226}     & \multicolumn{1}{l|}{1.265}     & \multicolumn{1}{l|}{-11.6} \\
\multicolumn{1}{|l|}{3.3}  & \multicolumn{1}{l|}{3.6}  & \multicolumn{1}{l|}{1.368}     & \multicolumn{1}{l|}{1.409}     & \multicolumn{1}{l|}{-9.3}   & \multicolumn{1}{l|}{1.213}     & \multicolumn{1}{l|}{1.250}     & \multicolumn{1}{l|}{-17.3} \\
\multicolumn{1}{|l|}{3.6}  & \multicolumn{1}{l|}{3.9}  & \multicolumn{1}{l|}{1.343}     & \multicolumn{1}{l|}{1.393}     & \multicolumn{1}{l|}{-12.0}  & \multicolumn{1}{l|}{1.204}     & \multicolumn{1}{l|}{1.248}     & \multicolumn{1}{l|}{-19.2} \\
\multicolumn{1}{|l|}{3.9}  & \multicolumn{1}{l|}{4.2}  & \multicolumn{1}{l|}{1.329}     & \multicolumn{1}{l|}{1.384}     & \multicolumn{1}{l|}{-12.6}  & \multicolumn{1}{l|}{1.195}     & \multicolumn{1}{l|}{1.242}     & \multicolumn{1}{l|}{-18.5} \\
\multicolumn{1}{|l|}{4.2}  & \multicolumn{1}{l|}{4.5}  & \multicolumn{1}{l|}{1.319}     & \multicolumn{1}{l|}{1.372}     & \multicolumn{1}{l|}{-13.4}  & \multicolumn{1}{l|}{1.191}     & \multicolumn{1}{l|}{1.237}     & \multicolumn{1}{l|}{-19.7} \\
\multicolumn{1}{|l|}{4.5}  & \multicolumn{1}{l|}{4.8}  & \multicolumn{1}{l|}{1.301}     & \multicolumn{1}{l|}{1.361}     & \multicolumn{1}{l|}{-15.1}  & \multicolumn{1}{l|}{1.184}     & \multicolumn{1}{l|}{1.229}     & \multicolumn{1}{l|}{-21.0} \\
\multicolumn{1}{|l|}{4.8}  & \multicolumn{1}{l|}{5.1}  & \multicolumn{1}{l|}{1.293}     & \multicolumn{1}{l|}{1.356}     & \multicolumn{1}{l|}{-16.7}  & \multicolumn{1}{l|}{1.182}     & \multicolumn{1}{l|}{1.228}     & \multicolumn{1}{l|}{-19.8} \\
\multicolumn{1}{|l|}{5.1}  & \multicolumn{1}{l|}{5.4}  & \multicolumn{1}{l|}{1.285}     & \multicolumn{1}{l|}{1.344}     & \multicolumn{1}{l|}{-16.7}  & \multicolumn{1}{l|}{1.177}     & \multicolumn{1}{l|}{1.222}     & \multicolumn{1}{l|}{-24.2} \\
\multicolumn{1}{|l|}{5.4}  & \multicolumn{1}{l|}{5.7}  & \multicolumn{1}{l|}{1.279}     & \multicolumn{1}{l|}{1.334}     & \multicolumn{1}{l|}{-17.3}  & \multicolumn{1}{l|}{1.171}     & \multicolumn{1}{l|}{1.218}     & \multicolumn{1}{l|}{-24.2} \\
\multicolumn{1}{|l|}{5.7}  & \multicolumn{1}{l|}{6.0}  & \multicolumn{1}{l|}{1.267}     & \multicolumn{1}{l|}{1.327}     & \multicolumn{1}{l|}{-19.2}  & \multicolumn{1}{l|}{1.168}     & \multicolumn{1}{l|}{1.212}     & \multicolumn{1}{l|}{-26.4} \\
\multicolumn{1}{|l|}{6.0}  & \multicolumn{1}{l|}{6.3}  & \multicolumn{1}{l|}{1.259}     & \multicolumn{1}{l|}{1.323}     & \multicolumn{1}{l|}{-21.4}  & \multicolumn{1}{l|}{1.164}     & \multicolumn{1}{l|}{1.211}     & \multicolumn{1}{l|}{-28.8} \\
\multicolumn{1}{|l|}{6.3}  & \multicolumn{1}{l|}{6.6}  & \multicolumn{1}{l|}{1.254}     & \multicolumn{1}{l|}{1.317}     & \multicolumn{1}{l|}{-21.7}  & \multicolumn{1}{l|}{1.160}     & \multicolumn{1}{l|}{1.208}     & \multicolumn{1}{l|}{-28.9} \\
\multicolumn{1}{|l|}{6.6}  & \multicolumn{1}{l|}{6.9}  & \multicolumn{1}{l|}{1.251}     & \multicolumn{1}{l|}{1.307}     & \multicolumn{1}{l|}{-21.5}  & \multicolumn{1}{l|}{1.155}     & \multicolumn{1}{l|}{1.201}     & \multicolumn{1}{l|}{-33.2} \\
\multicolumn{1}{|l|}{6.9}  & \multicolumn{1}{l|}{7.2}  & \multicolumn{1}{l|}{1.240}     & \multicolumn{1}{l|}{1.301}     & \multicolumn{1}{l|}{-24.5}  & \multicolumn{1}{l|}{1.154}     & \multicolumn{1}{l|}{1.201}     & \multicolumn{1}{l|}{-32.5} \\
\multicolumn{1}{|l|}{7.2}  & \multicolumn{1}{l|}{7.5}  & \multicolumn{1}{l|}{1.233}     & \multicolumn{1}{l|}{1.298}     & \multicolumn{1}{l|}{-26.8}  & \multicolumn{1}{l|}{1.147}     & \multicolumn{1}{l|}{1.198}     & \multicolumn{1}{l|}{-38.0} \\
\multicolumn{1}{|l|}{7.5}  & \multicolumn{1}{l|}{7.8}  & \multicolumn{1}{l|}{1.236}     & \multicolumn{1}{l|}{1.293}     & \multicolumn{1}{l|}{-23.1}  & \multicolumn{1}{l|}{1.155}     & \multicolumn{1}{l|}{1.195}     & \multicolumn{1}{l|}{-29.2} \\
\multicolumn{1}{|l|}{7.8}  & \multicolumn{1}{l|}{8.1}  & \multicolumn{1}{l|}{1.232}     & \multicolumn{1}{l|}{1.285}     & \multicolumn{1}{l|}{-24.3}  & \multicolumn{1}{l|}{1.149}     & \multicolumn{1}{l|}{1.192}     & \multicolumn{1}{l|}{-36.9} \\
\multicolumn{1}{|l|}{8.1}  & \multicolumn{1}{l|}{8.4}  & \multicolumn{1}{l|}{1.223}     & \multicolumn{1}{l|}{1.280}     & \multicolumn{1}{l|}{-27.2}  & \multicolumn{1}{l|}{1.142}     & \multicolumn{1}{l|}{1.189}     & \multicolumn{1}{l|}{-37.2} \\
\multicolumn{1}{|l|}{8.4}  & \multicolumn{1}{l|}{8.7}  & \multicolumn{1}{l|}{1.225}     & \multicolumn{1}{l|}{1.278}     & \multicolumn{1}{l|}{-24.6}  & \multicolumn{1}{l|}{1.145}     & \multicolumn{1}{l|}{1.187}     & \multicolumn{1}{l|}{-36.8} \\
\multicolumn{1}{|l|}{8.7}  & \multicolumn{1}{l|}{9.0}  & \multicolumn{1}{l|}{1.221}     & \multicolumn{1}{l|}{1.273}     & \multicolumn{1}{l|}{-25.7}  & \multicolumn{1}{l|}{1.143}     & \multicolumn{1}{l|}{1.184}     & \multicolumn{1}{l|}{-36.8} \\
\multicolumn{1}{|l|}{9.0}  & \multicolumn{1}{l|}{9.3}  & \multicolumn{1}{l|}{1.216}     & \multicolumn{1}{l|}{1.268}     & \multicolumn{1}{l|}{-26.3}  & \multicolumn{1}{l|}{1.145}     & \multicolumn{1}{l|}{1.182}     & \multicolumn{1}{l|}{-33.3} \\
\multicolumn{1}{|l|}{9.3}  & \multicolumn{1}{l|}{9.6}  & \multicolumn{1}{l|}{1.211}     & \multicolumn{1}{l|}{1.264}     & \multicolumn{1}{l|}{-29.8}  & \multicolumn{1}{l|}{1.135}     & \multicolumn{1}{l|}{1.179}     & \multicolumn{1}{l|}{-47.3} \\
\multicolumn{1}{|l|}{9.6}  & \multicolumn{1}{l|}{9.9}  & \multicolumn{1}{l|}{1.206}     & \multicolumn{1}{l|}{1.261}     & \multicolumn{1}{l|}{-30.0}  & \multicolumn{1}{l|}{1.137}     & \multicolumn{1}{l|}{1.179}     & \multicolumn{1}{l|}{-40.4} \\
\multicolumn{1}{|l|}{9.9}  & \multicolumn{1}{l|}{10.2} & \multicolumn{1}{l|}{1.209}     & \multicolumn{1}{l|}{1.255}     & \multicolumn{1}{l|}{-26.7}  & \multicolumn{1}{l|}{1.141}     & \multicolumn{1}{l|}{1.177}     & \multicolumn{1}{l|}{-36.2} \\
\multicolumn{1}{|l|}{10.2} & \multicolumn{1}{l|}{10.5} & \multicolumn{1}{l|}{1.206}     & \multicolumn{1}{l|}{1.254}     & \multicolumn{1}{l|}{-29.1}  & \multicolumn{1}{l|}{1.139}     & \multicolumn{1}{l|}{1.175}     & \multicolumn{1}{l|}{-38.4} \\
\multicolumn{1}{|l|}{10.5} & \multicolumn{1}{l|}{10.8} & \multicolumn{1}{l|}{1.201}     & \multicolumn{1}{l|}{1.249}     & \multicolumn{1}{l|}{-30.7}  & \multicolumn{1}{l|}{1.134}     & \multicolumn{1}{l|}{1.173}     & \multicolumn{1}{l|}{-40.6} \\
\multicolumn{1}{|l|}{10.8} & \multicolumn{1}{l|}{11.1} & \multicolumn{1}{l|}{1.201}     & \multicolumn{1}{l|}{1.244}     & \multicolumn{1}{l|}{-29.0}  & \multicolumn{1}{l|}{1.135}     & \multicolumn{1}{l|}{1.171}     & \multicolumn{1}{l|}{-40.0} \\
\multicolumn{1}{|l|}{11.1} & \multicolumn{1}{l|}{11.4} & \multicolumn{1}{l|}{1.195}     & \multicolumn{1}{l|}{1.242}     & \multicolumn{1}{l|}{-31.8}  & \multicolumn{1}{l|}{1.134}     & \multicolumn{1}{l|}{1.169}     & \multicolumn{1}{l|}{-39.7} \\
\multicolumn{1}{|l|}{11.4} & \multicolumn{1}{l|}{11.7} & \multicolumn{1}{l|}{1.199}     & \multicolumn{1}{l|}{1.240}     & \multicolumn{1}{l|}{-28.9}  & \multicolumn{1}{l|}{1.139}     & \multicolumn{1}{l|}{1.168}     & \multicolumn{1}{l|}{-31.4} \\
\multicolumn{1}{|l|}{11.7} & \multicolumn{1}{l|}{12.0} & \multicolumn{1}{l|}{1.197}     & \multicolumn{1}{l|}{1.239}     & \multicolumn{1}{l|}{-29.3}  & \multicolumn{1}{l|}{1.137}     & \multicolumn{1}{l|}{1.168}     & \multicolumn{1}{l|}{-35.9} \\
\multicolumn{1}{|l|}{12.0} & \multicolumn{1}{l|}{12.3} & \multicolumn{1}{l|}{1.189}     & \multicolumn{1}{l|}{1.232}     & \multicolumn{1}{l|}{-28.9}  & \multicolumn{1}{l|}{1.132}     & \multicolumn{1}{l|}{1.164}     & \multicolumn{1}{l|}{-35.3} \\
\multicolumn{1}{|l|}{12.3} & \multicolumn{1}{l|}{12.6} & \multicolumn{1}{l|}{1.187}     & \multicolumn{1}{l|}{1.230}     & \multicolumn{1}{l|}{-34.7}  & \multicolumn{1}{l|}{1.130}     & \multicolumn{1}{l|}{1.163}     & \multicolumn{1}{l|}{-43.0} \\
\multicolumn{1}{|l|}{12.6} & \multicolumn{1}{l|}{12.9} & \multicolumn{1}{l|}{1.192}     & \multicolumn{1}{l|}{1.228}     & \multicolumn{1}{l|}{-28.1}  & \multicolumn{1}{l|}{1.133}     & \multicolumn{1}{l|}{1.162}     & \multicolumn{1}{l|}{-37.7} \\
\multicolumn{1}{|l|}{12.9} & \multicolumn{1}{l|}{13.2} & \multicolumn{1}{l|}{1.185}     & \multicolumn{1}{l|}{1.226}     & \multicolumn{1}{l|}{-32.1}  & \multicolumn{1}{l|}{1.133}     & \multicolumn{1}{l|}{1.161}     & \multicolumn{1}{l|}{-35.4} \\
\multicolumn{1}{|l|}{13.2} & \multicolumn{1}{l|}{13.5} & \multicolumn{1}{l|}{1.184}     & \multicolumn{1}{l|}{1.224}     & \multicolumn{1}{l|}{-34.3}  & \multicolumn{1}{l|}{1.127}     & \multicolumn{1}{l|}{1.160}     & \multicolumn{1}{l|}{-48.8} \\
\multicolumn{1}{|l|}{13.5} & \multicolumn{1}{l|}{13.8} & \multicolumn{1}{l|}{1.183}     & \multicolumn{1}{l|}{1.220}     & \multicolumn{1}{l|}{-32.7}  & \multicolumn{1}{l|}{1.132}     & \multicolumn{1}{l|}{1.158}     & \multicolumn{1}{l|}{-38.5} \\
\multicolumn{1}{|l|}{13.8} & \multicolumn{1}{l|}{14.1} & \multicolumn{1}{l|}{1.185}     & \multicolumn{1}{l|}{1.217}     & \multicolumn{1}{l|}{-28.6}  & \multicolumn{1}{l|}{1.133}     & \multicolumn{1}{l|}{1.157}     & \multicolumn{1}{l|}{-34.6} \\
\multicolumn{1}{|l|}{14.1} & \multicolumn{1}{l|}{14.4} & \multicolumn{1}{l|}{1.179}     & \multicolumn{1}{l|}{1.216}     & \multicolumn{1}{l|}{-35.4}  & \multicolumn{1}{l|}{1.123}     & \multicolumn{1}{l|}{1.156}     & \multicolumn{1}{l|}{-52.8} \\
\multicolumn{1}{|l|}{14.4} & \multicolumn{1}{l|}{14.7} & \multicolumn{1}{l|}{1.181}     & \multicolumn{1}{l|}{1.213}     & \multicolumn{1}{l|}{-31.1}  & \multicolumn{1}{l|}{1.127}     & \multicolumn{1}{l|}{1.154}     & \multicolumn{1}{l|}{-43.5} \\
\multicolumn{1}{|l|}{14.7} & \multicolumn{1}{l|}{15.0} & \multicolumn{1}{l|}{1.180}     & \multicolumn{1}{l|}{1.210}     & \multicolumn{1}{l|}{-31.4}  & \multicolumn{1}{l|}{1.130}     & \multicolumn{1}{l|}{1.153}     & \multicolumn{1}{l|}{-39.2} \\ \hline
\end{tabular}
\caption{This table summarises the detour for commuting trips ($\delta_c$) and for typical network values ($\delta_n$) within distance ranges of crow-fly distance. It shows how they significantly differentiate, especially as crow fly distance grows (column $d$). We test the significance in the difference of the observed means for the weighted detour compared with regular network values. The results show that the observed difference are strongly significant, especially as the crow fly distance between locations increases.}
\label{detour_sign_tab}
\end{table}

\end{document}